\documentclass[aps,prb,reprint,superscriptaddress,longbibliography]{revtex4-2}
\usepackage{graphicx}
\usepackage{amsmath}
\usepackage{amssymb}
\usepackage{bm}
\usepackage[utf8]{inputenc}

\newcommand{\ctext}[1]{\raise0.2ex\hbox{\textcircled{\scriptsize{#1}}}}
\begin{document}

\title{Symmetry-Selective Shift Current in Polar Multidomain Superlattices}

\author{Tomoaki Kameda}
\email{t.kameda@kwansei.ac.jp}
\affiliation{Department of Nanotechnology for Sustainable Energy,
  School of Science and Technology, Kwansei Gakuin University,
  Gakuen-Uegahara 1, Sanda 669-1330, Japan}
\affiliation{Research Center for Materials Nanoarchitectonics (MANA),
  National Institute for Materials Science (NIMS), Namiki
  1-1, Tsukuba 305-0044, Japan}
\author{Kosuke Nagashio}
\affiliation{Department of Materials Engineering, The University of Tokyo,
  Bunkyo, Japan}
\author{Katsunori Wakabayashi}
\email{WAKABAYASHI.Katsunori@nims.go.jp}
\affiliation{Research Center for Materials Nanoarchitectonics (MANA),
  National Institute for Materials Science (NIMS), Namiki
  1-1, Tsukuba 305-0044, Japan}
\affiliation{Department of Nanotechnology for Sustainable Energy,
  School of Science and Technology, Kwansei Gakuin University,
  Gakuen-Uegahara 1, Sanda 669-1330, Japan}
\affiliation{Center for Spintronics Research Network (CSRN), The University of Osaka, Toyonaka 560-8531, Japan}

\begin{abstract}
Periodic polar multidomain superlattices can act as symmetry-selective elements for shift-current generation. We demonstrate this using a two-dimensional Su--Schrieffer--Heeger model with prescribed multidomain configurations. The distinct full-supercell symmetries of periodic head-to-head and head-to-tail arrangements impose different constraints on the shift-current tensor and thereby select complementary nonlinear-response channels, as confirmed by numerical calculations. A direct comparison with equal-area averages of the constituent single-domain spectra shows that an independent-domain average does not reproduce the detailed multidomain spectra. A bond-resolved contribution analysis further clarifies the real-space origin of the multidomain response.
Widening each domain from one to ten primitive cells leaves the selection rules intact. On the photon-energy grid used here, the allowed head-to-tail response changes sign and grows by about an order of magnitude. These results provide a theoretical framework for the symmetry-based design of anisotropic optical rectification in polar multidomain superlattices.
\end{abstract}

\maketitle

\section{Introduction}
The bulk photovoltaic effect (BPVE) provides a distinct route to dc photocurrent generation in homogeneous noncentrosymmetric crystals without an external bias~\cite{Glass1974,KrautBaltz1979,vonBaltzKraut1981,SturmanFridkin1992,Spanier2016,DaiRappe2023,Shyam2026Review}.
This intrinsic bulk response requires neither a p--n junction nor a built-in junction field.
It therefore differs from conventional junction-based conversion, in which photogenerated carriers are separated by the built-in field and the maximum efficiency is set by the detailed-balance limit~\cite{ShockleyQueisser1961}, and from the surface-field mechanisms proposed in early ferroelectric photovoltaic studies~\cite{Chynoweth1956}.
One microscopic mechanism of the BPVE is shift current, a second-order nonlinear optical response associated with the real-space displacement of electronic wave packets during interband optical transitions~\cite{SipeGhahramani1993,AversaSipe1995,SipeShkrebtii2000,NastosSipe2010,YoungRappe2012,YoungZhengRappe2012,Tan2016,Fregoso2017,MorimotoNagaosa2016,Ahn2020,IbanezAzpiroz2018,TanRappe2016TopologicalBPVE,YoshidaMurakami2025Gapless,Avdoshkin2025}.
Shift current has attracted considerable attention as an intrinsic photocurrent mechanism that may overcome some limitations of conventional photovoltaics, including band-gap-related voltage constraints and scattering-limited carrier transport~\cite{ShockleyQueisser1961,YoungRappe2012,Cook2017,Pusch2023,Feng2025HighEfficiency}.

\begin{figure*}
  \centering
  \includegraphics[width=170mm]{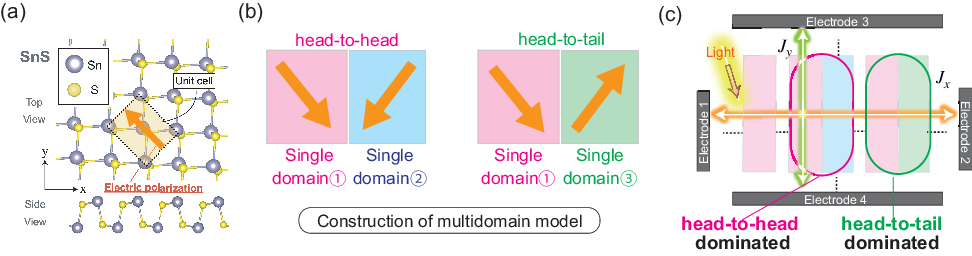}
  \caption{Conceptual overview of symmetry-selective shift current in polar multidomain superlattices.
(a) Ferroelectric SnS provides a representative polar two-dimensional system in which light irradiation induces shift current through the bulk photovoltaic effect.
The SnS structure motivates the study; its atomic geometry is not identified with the model lattice in (b).
(b) Periodic head-to-head and head-to-tail arrangements are modeled by combining polar single-domain building blocks in a minimal two-dimensional SSH model.
(c) Conceptual design principle for anisotropic optical rectification in a mixed multidomain structure, motivated by the dominant response channels.
Head-to-head-dominated regions favor a $y$-direction response, whereas head-to-tail-dominated regions favor an $x$-direction response.
}\label{fig:overview}
\end{figure*}

Two-dimensional (2D) ferroelectrics provide a promising platform for shift current because their switchable polar order breaks inversion symmetry~\cite{Fei2016,Kruse2023Screening}.
In particular, group-IV monochalcogenide monolayers such as GeS and SnS combine ferroelectric order~\cite{Fei2016,Higashitarumizu2020SnS,WangQian2017} with large or tunable shift-current responses~\cite{Cook2017,Kushnir2017GeS,Kaner2020StrainGeSSnS,EsteveParedes2025}.
A representative SnS structure is shown in Fig.~\ref{fig:overview}(a).
Robust in-plane ferroelectricity in related monochalcogenides and the tunability of their polar order have also been reported~\cite{Chang2016SnTe,BarrazaLopez2018}.
Beyond this family, sizable bulk photovoltaic responses have been reported in elemental 2D ferroelectrics~\cite{Qian2023Elemental}, in van der Waals ferroelectrics such as CuInP$_2$S$_6$~\cite{Li2021CIPS}, and in rhombohedrally stacked or atomically thin transition-metal dichalcogenides~\cite{Yang2022MoS2,Dong2023MoS2,Ramos2024ReS2}.
The attainable photovoltaic efficiency of 2D shift-current materials has been assessed systematically~\cite{Sauer2023}, and routes to enhancing the response have been proposed~\cite{Chen2024Virtual}.

For SnS itself, effective tight-binding models derived from first-principles calculations reproduce the essential features of its monolayer shift-current spectrum and identify the interband transitions that dominate it~\cite{Kusunoki2026SnSTB}.
Experiments on in-plane ferroelectric SnS have demonstrated shift-current photovoltaics and examined BPVE responses in single-domain and domain-structured crystals~\cite{Chang2023SnS,Nanae2024SnS}.
In multidomain SnS, photocurrent measurements further show that the domain-wall contribution to the photovoltaic response depends on the arrangement of the domain walls relative to the electrodes~\cite{Nanae2026DomainSnS}.
Ferroelectric domain walls can host electronic structures and conductivities distinct from those of the adjoining domains~\cite{Seidel2009,Catalan2012,Meier2015}, and domain-wall potential steps and photoconductivity can control photovoltaic responses in BiFeO$_3$~\cite{Yang2010,Bhatnagar2013}.
First-principles supercell calculations describe ferroelectric domain
walls in perovskite oxides~\cite{Meyer2002PbTiO3}.
For group-IV monochalcogenide monolayers, first-principles calculations
also show that neutral domain walls modify the electronic structure and
linear optical absorption~\cite{Petralanda2023DomainWalls}.
Localized BPVE responses at van der Waals grain boundaries, polar textures, and polar interfaces further demonstrate that spatial inhomogeneity can reshape photocurrent generation~\cite{Zhou2024GrainBoundary,Agarwal2025Vortices,Akamatsu2021}.
These observations motivate a theoretical description that goes beyond spatially uniform bulk models and includes multidomain structures and their interfaces explicitly.

In this paper, we investigate how multidomain geometry modifies shift-current responses using the Su--Schrieffer--Heeger (SSH) model on a 2D square lattice~\cite{Su1979,LiuWakabayashi2017,Obana2019SSH,Benalcazar2017Science,Benalcazar2017PRB}.
This study is motivated in particular by the anisotropic domain-wall photovoltaic response reported for multidomain SnS~\cite{Nanae2026DomainSnS}.
Our aim is to examine how prescribed polar arrangements constrain the in-plane response tensor. The square-lattice model is not a material-specific Hamiltonian for SnS.
The staggered on-site potentials of this model prescribe inversion-symmetry-broken polar configurations rather than dynamically switchable ferroelectric states, so that it serves as an effective polar model of ferroelectric multidomain systems.
The nonlinear conductivity is a third-rank tensor, so crystal symmetry dictates which of its components may be finite.
This constraint has been used to analyze and manipulate nonlinear optical responses in two-dimensional materials~\cite{Habara2022NbSe2,HabaraWakabayashi2023,Zhou2020NHE}.

As shown in Fig.~\ref{fig:overview}(b), we focus on periodic head-to-head and head-to-tail arrangements constructed from the same polar building blocks, and we show that they select different nonlinear-response channels through their distinct symmetries.
This component selection is governed by the symmetry of the full multidomain geometry.
The magnitude and spectral profile of the allowed responses, however, cannot be inferred from a simple area average of the constituent single-domain spectra.
Instead, they must be determined from the electronic structure, optical transition matrix elements, and shift vectors of the full multidomain supercell.
We also introduce a bond-resolved decomposition to analyze the real-space origin of the multidomain response.
Together, these results establish a conceptual design principle for anisotropic optical rectification based on the dominant response components selected by multidomain geometry.
This design principle is schematically illustrated in Fig.~\ref{fig:overview}(c).

Section~\ref{sec:model-polarization} introduces the single-domain and multidomain 2D SSH models and classifies the domains by their bulk electronic polarization. Section~\ref{sec:symmetry-constraints} derives symmetry constraints on the shift-current tensor, and Sec.~\ref{sec:single-domain-shift} summarizes the formalism and establishes the single-domain baseline. Section~\ref{sec:multidomain-shift} analyzes the periodic head-to-head and head-to-tail arrangements, while Sec.~\ref{sec:real-space} provides a real-space interpretation of their responses. Section~\ref{sec_conclusion} concludes the paper. Appendixes~\ref{sec:appendixA} and~\ref{sec:appendixB} present additional tensor components and the polarization-angle dependence, respectively. Appendix~\ref{sec:appendixC} examines wider domains and their bond-resolved responses.

\section{Model and Polarization Structure}
\label{sec:model-polarization}
We first define the single-domain 2D SSH model and its electronic polarization, and then construct the multidomain supercells used below.

\subsection{Single-domain 2D SSH model}
The single-domain 2D SSH model~\cite{LiuWakabayashi2017,Obana2019SSH} serves as the bulk building block of the multidomain structures studied below.
Throughout the numerical calculations, energies are expressed in units of a
reference model energy $E_{\rm ref}$ and lengths in units of the
lattice constant $a$, so that $E_{\rm ref}=1$ and $a=1$ numerically.
All energy parameters quoted below, including $\gamma$,
$\gamma^{\prime}$, and $\Delta_{\alpha}$, therefore denote their values relative to
$E_{\rm ref}$.

\begin{figure}
  \centering
  \includegraphics[width=85mm]{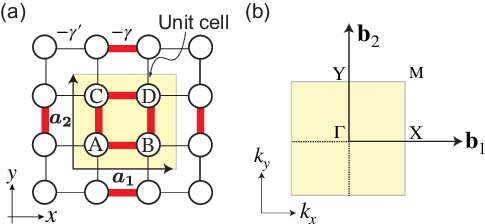}
  \caption{
    (a) Schematic of the SSH model on a square lattice. Thick (red) and
    thin (black) bonds represent the intracell and intercell hoppings,
    $-\gamma$ and $-\gamma^{\prime}$, respectively.
    The primitive translation vectors are $\bm{a}_1=(a,0)$
    and $\bm{a}_2=(0,a)$. The yellow-shaded square indicates
    the four-site unit cell.
    (b) Corresponding first Brillouin zone (BZ) with high-symmetry points
    $\Gamma$, $\mathrm{X}$, $\mathrm{M}$, and $\mathrm{Y}$.
    The reciprocal lattice vectors are $\bm{b}_1=(\frac{2\pi}{a}, 0)$ and $\bm{b}_2=(0, \frac{2\pi}{a})$.
  }\label{fig:2d-ssh-lattice}
\end{figure}
Figure~\ref{fig:2d-ssh-lattice}(a) defines the four-site square-lattice unit cell, and Fig.~\ref{fig:2d-ssh-lattice}(b) shows the corresponding first Brillouin zone (BZ). Each site hosts a single electronic orbital, and $\gamma$ and $\gamma^{\prime}$ are positive real hopping amplitudes.

Monolayer SnS has a puckered structure with atoms at different out-of-plane heights and a rectangular primitive cell containing two Sn and two S atoms~\cite{Fei2016,Kusunoki2026SnSTB}.
The present planar square lattice omits this thickness-direction structure and does not assign its four orbitals to specific Sn or S atomic orbitals.
It is a minimal polar model, not a downfolding of the SnS electronic structure.
For comparison, a material-specific description based on first-principles Wannier functions uses twelve $p$ orbitals per primitive cell~\cite{Kusunoki2026SnSTB}.

For the polar configurations studied below, we choose $\gamma=1.7$ and $\gamma^{\prime}=0.3$, with one on-site potential $-2$ and the other three $2/3$ in each constituent four-site cell, as representative model parameters.
These values give unequal hopping amplitudes and an inversion-symmetry-broken on-site pattern. They are not fitted to the band structure or optical spectrum of SnS.
The resulting band dispersions and response spectra therefore characterize this model parameter set, and no quantitative agreement with SnS is claimed.

The eigenvalue equation of the 2D SSH model on a square lattice is written as
\begin{equation}
  \hat{H}(\mathbf{k})|u_n(\mathbf{k})\rangle = E_{n}(\mathbf{k})|u_n(\mathbf{k})\rangle,
\end{equation}
where $\mathbf{k}=(k_x,k_y)$ is the wave vector,
$E_n(\mathbf{k})$ is the eigenenergy, and $n=1,2,3,4$ denotes the band index.
The eigenvector is defined by
\[
\lvert u_n(\mathbf{k})\rangle
  =\bigl(\psi_{n,A}(\mathbf{k}),\,
         \psi_{n,B}(\mathbf{k}),\,
         \psi_{n,C}(\mathbf{k}),\,
         \psi_{n,D}(\mathbf{k})\bigr)^{T},
\]
where $(\cdots)^{T}$ denotes the transpose of a vector.
Here, $\psi_{n,\alpha}(\mathbf{k})$ ($\alpha=A,B,C,D$) is the amplitude on
site $\alpha$ in the $n$th energy band at $\mathbf{k}$.

The single-domain (SD) Bloch Hamiltonian $\hat{H}_{\rm{SD}}(\mathbf{k})$ is explicitly written as
\begin{equation}
\hat{H}_{\rm{SD}}(\mathbf{k})=
\begin{pmatrix}
\Delta_{\rm A}                & \rho_x(k_x)    & \rho_y(k_y)    & 0 \\
\rho_x^{*}(k_x)  & \Delta_{\rm B}              & 0              & \rho_y(k_y) \\
\rho_y^{*}(k_y)  & 0              & \Delta_{\rm C}              & \rho_x(k_x) \\
0                & \rho_y^{*}(k_y)& \rho_x^{*}(k_x)& \Delta_{\rm D}
\end{pmatrix},
\label{eq:HkSD}
\end{equation}
where
\begin{equation}
\rho_{\ell}(k_{\ell})=-\gamma e^{-i k_{\ell}\frac{a}{2}}-\gamma^{\prime} e^{ik_\ell \frac{a}{2}},
\qquad (\ell=x,y).
\end{equation}
Here, we introduce staggered on-site potentials $\Delta_{\alpha}$ ($\alpha={\rm A,B,C,D}$) to break inversion symmetry and induce a finite electronic polarization in the bulk. The resulting asymmetric charge distribution within the unit cell is what produces the polarization.
The present electronic tight-binding model does not include ionic polarization, structural relaxation, a double-well energy landscape, a switching path or coercive field, depolarization fields, charge screening, or self-consistent electrostatics. It should therefore be viewed as a prescribed polar model rather than a microscopic switching model.
\begin{figure}
  \centering
  \includegraphics[width=85mm]{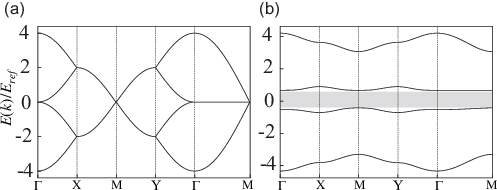}
  \caption{
    Energy band structures of the 2D SSH model for (a) $\gamma = \gamma^{\prime} = 1$, $\Delta_{\rm A}=\Delta_{\rm B}=\Delta_{\rm C}=\Delta_{\rm D} = 0$, and (b) $\gamma = 1.7$, $\gamma^{\prime} = 0.3$, $\Delta_{\rm A}=\Delta_{\rm C}=\Delta_{\rm D} = \frac{2}{3}$, and $\Delta_{\rm B}=-2$. Modulation of hoppings and on-site terms resolves the degeneracies and induces finite band gaps.
    At half filling the two lower bands are occupied, so bands $1$ and $2$ are
    the valence bands and bands $3$ and $4$ the conduction bands. The gray
    region marks the central gap between the second and third bands, which is
    the optical gap relevant to the shift-current response.
    Energies are given in units of the reference energy $E_{\rm ref}$ used
    throughout.
  }\label{fig:2d-ssh-energy_band}
\end{figure}
Figure~\ref{fig:2d-ssh-energy_band} compares the symmetric case with the case of broken inversion symmetry. In Fig.~\ref{fig:2d-ssh-energy_band}(a), where $\gamma=\gamma^{\prime}$ and the on-site potentials vanish ($\Delta_{\alpha}=0$), all high-symmetry band crossings remain gapless. By contrast, in Fig.~\ref{fig:2d-ssh-energy_band}(b), $\gamma \neq \gamma^{\prime}$ and finite staggered on-site potentials ($\Delta_{\alpha} \neq 0$) break inversion symmetry and open band gaps at high-symmetry points in the Brillouin zone.

\subsection{Electronic polarization and classification of single domains}
We classify the single domains by the electronic polarization induced by their staggered on-site potentials. This quantity allows the interfaces formed by domains with different polarization directions to be classified in turn.
In tight-binding lattices the electronic polarization is expressed through the Berry phases of the occupied Bloch states~\cite{Zak1989}. When symmetry quantizes these phases, the quantized value is the topological invariant of the Su--Schrieffer--Heeger model~\cite{LiuWakabayashi2017,Obana2019SSH,Kameda2019Zak}.

In a one-dimensional (1D) crystalline solid, the electronic polarization $P$ can be defined as the dipole moment per unit-cell volume and expressed as~\cite{KingSmithVanderbilt1993,Resta1994}
\begin{align}
P
= \sum_{n}^{\text{occ}}P^{n}
= -\frac{e}{V}\sum_{n}^{\text{occ}} r_n
= -\frac{e}{2\pi}\sum_{n}^{\text{occ}}\int_{\text{BZ}} dk \, A_n(k),
\end{align}
with
\begin{align}
r_n
= \frac{V}{2\pi}\int_{\text{BZ}} dk \, A_n(k),
\end{align}
where $r_n$ is the Wannier center of the $n$th occupied band, i.e., the expectation value of the position operator for the corresponding Wannier function~\cite{Blount1962BandTheory,Marzari1997Wannier,Hamai2023Wannier,Hamai2026MLWF}.
We define the Berry connection of the $n$th energy band as
\begin{align}
A_n(k)
= i\langle u_n(k)|\partial_k u_n(k)\rangle.
\end{align}
With this convention, the position operator projected onto an isolated Bloch band is represented by $i\partial_k + A_n(k)$ in the Bloch representation~\cite{Blount1962BandTheory}. Here, $e>0$ denotes the elementary charge (the electron charge is $-e$), $V$ is the unit-cell volume, which reduces to the lattice constant in one dimension, and the sum runs over all occupied bands. The quantity $P^{n}$ is the contribution of the $n$th band to the total polarization.

In a 2D crystalline system, the electronic polarization is a vector
$\mathbf{P}^{\rm el}=(P_x^{\rm el},P_y^{\rm el})$. The corresponding Berry
connection of the $n$th band is
\[
\mathbf{A}_n(\mathbf{k})
=i\langle u_n(\mathbf{k})|\nabla_{\mathbf{k}}u_n(\mathbf{k})\rangle.
\]
For the square lattice considered here, we define the contribution of the
$n$th band to the electronic polarization along the $i$ direction as
\begin{align}
P_i^n
=-\frac{e}{(2\pi)^2}
\int_{\mathcal{C}_j}dk_j
\oint_{\mathcal{C}_i(k_j)}dk_i\,
\mathbf{A}_n(k_i,k_j)\cdot\mathbf{n}_i,
\end{align}
and the total electronic polarization is
\begin{align}
P_i^{\rm el}=\sum_{n\in\mathrm{occ}}P_i^n.
\label{eq:Pel_sum}
\end{align}
Equivalently, introducing the non-Abelian Berry connection of the occupied
subspace,
\begin{align}
\mathcal{A}^{mn}_i(\mathbf{k})
=i\langle u_m(\mathbf{k})|\partial_{k_i}u_n(\mathbf{k})\rangle
\qquad (m,n\in\mathrm{occ}),
\end{align}
Eq.~\eqref{eq:Pel_sum} can be written as
\begin{align}
P_i^{\rm el}
=-\frac{e}{(2\pi)^2}\int_{\rm BZ}d^2k\;
\mathrm{Tr}\,\mathcal{A}_i(\mathbf{k}),
\label{eq:Pel_trace}
\end{align}
where the trace runs over the occupied bands.

Under a smooth unitary mixing of
the occupied states,
$|u_n(\mathbf{k})\rangle\rightarrow\sum_{m\in\mathrm{occ}}
|u_m(\mathbf{k})\rangle U_{mn}(\mathbf{k})$,
the non-Abelian Berry connection transforms as
$\mathcal{A}_i\rightarrow U^{\dagger}\mathcal{A}_i U
+iU^{\dagger}\partial_{k_i}U$,
so that $\mathrm{Tr}\,\mathcal{A}_i$ acquires the additional term
$i\,\partial_{k_i}\ln\det U$. Because this term integrates to an integer
multiple of $2\pi$ along a closed reciprocal-space loop, the trace formulation
is invariant under smooth unitary mixing of the occupied subspace up to the
usual polarization quantum associated with the winding of the gauge
transformation, given in Eq.~\eqref{eq:pol-quantum} below. The individual band
contributions $P_i^n$, by contrast, are not separately well defined once the
occupied bands mix or become degenerate. We therefore use the summed quantity
of Eq.~\eqref{eq:Pel_trace} throughout, and the band-resolved decomposition of
Eq.~\eqref{eq:Pel_sum} only as bookkeeping.
Here, $(i,j)=(x,y)$ or $(y,x)$, so the definitions of $P_x^{\rm el}$ and
$P_y^{\rm el}$ involve different Berry-connection components and do not imply
$P_x^{\rm el}=P_y^{\rm el}$. At fixed $k_j$,
$\mathcal{C}_i(k_j)$ is a primitive reciprocal-space loop along the $i$
direction whose endpoints are equivalent in the BZ. The path $\mathcal{C}_j$
spans one primitive reciprocal-space interval along the $j$ direction, and
$\mathbf{n}_i$ is the unit vector along the $i$ direction. The present
four-band Hamiltonian is spinless, and we evaluate it at half filling with the
two lower bands occupied. Hence, no additional spin-degeneracy factor is
included.

As in the modern theory of polarization, the electronic polarization is
defined modulo a polarization quantum,
\begin{align}
\mathbf{P}^{\rm el}
\equiv \mathbf{P}^{\rm el}+\frac{e\mathbf{R}}{A_{\rm cell}},
\qquad A_{\rm cell}=a^2,
\label{eq:pol-quantum}
\end{align}
where $\mathbf{R}$ is a lattice vector, so that $\mathbf{P}^{\rm el}$ carries
the dimension of charge per unit length, as appropriate for a 2D system. For
the three single-domain configurations considered below, we use the same
unit-cell convention and evaluate the Berry-connection integral of
Eq.~\eqref{eq:Pel_trace} on a common polarization branch, so that the reported
values may be compared with one another. We quote $\mathbf{P}^{\rm el}$ in
units of $e/(2\pi a)$, with $a=1$ in the numerical calculations.

\begin{figure}
  \centering
  \includegraphics[width=85mm]{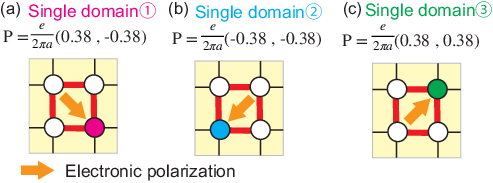}
  \caption{
    Schematic of the electronic polarization in the single-domain 2D SSH model.
    The arrows indicate the electronic polarizations evaluated from the occupied-band Berry-connection integrals on the common branch used for all three configurations.
    (a) Single domain \ctext{1}, where the on-site potentials are set to $\Delta_{\rm A}=\Delta_{\rm C}=\Delta_{\rm D}=\frac{2}{3}$ and $\Delta_{\rm B}=-2$, yielding the electronic polarization $\mathbf{P}^{\rm el}=\frac{e}{2\pi a}(0.38,\,-0.38)$.
    (b) Single domain \ctext{2}, where the on-site potentials are set to $\Delta_{\rm B}=\Delta_{\rm C}=\Delta_{\rm D} = \frac{2}{3}$ and $\Delta_{\rm A}=-2$, yielding the electronic polarization $\mathbf{P}^{\rm el}=\frac{e}{2\pi a}(-0.38,\,-0.38)$.
    (c) Single domain \ctext{3}, where the on-site potentials are set to $\Delta_{\rm A}=\Delta_{\rm B}=\Delta_{\rm C} = \frac{2}{3}$ and $\Delta_{\rm D}=-2$, yielding the electronic polarization $\mathbf{P}^{\rm el}=\frac{e}{2\pi a}(0.38,\,0.38)$.
    The hopping parameters are fixed at $\gamma=1.7$ and $\gamma^{\prime}=0.3$ in all panels (a)--(c).
  }\label{fig:single-domain-polarization}
\end{figure}
Figures~\ref{fig:single-domain-polarization}(a)--(c) show the three single-domain configurations, denoted $\ctext{1}$, $\ctext{2}$, and $\ctext{3}$, together with the electronic polarizations obtained from Eq.~\eqref{eq:Pel_trace} on a common branch.

This classification provides the basis for distinguishing the two
multidomain supercells considered below, constructed from single domains
$\ctext{1}$ and $\ctext{2}$ and from single domains $\ctext{1}$ and
$\ctext{3}$, respectively. Their periodic interface arrangements are defined
explicitly in the following subsection.

\subsection{Construction of multidomain structures}
We construct a doubled-unit-cell multidomain model by joining regions with different electronic polarization directions.
In the supercells studied in the main text, each constituent domain is one primitive cell wide in the $x$
direction. The doubled cell therefore represents a periodic polar
superlattice rather than an isolated domain wall. Appendix~\ref{sec:appendixC}
extends each constituent domain to five and ten primitive cells along $x$.

\begin{figure}
  \centering
  \includegraphics[width=84mm]{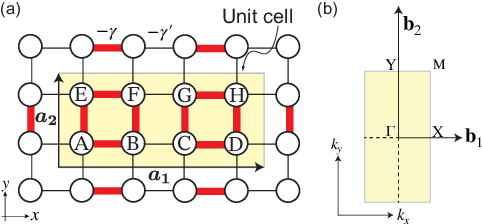}
  \caption{
    (a) Schematic of the multidomain model on its doubled unit cell. Thick (red) and
    thin (black) bonds represent the intracell and intercell hoppings,
    $-\gamma$ and $-\gamma^{\prime}$, respectively.
    The primitive translation vectors are $\bm{a}_1=(2a,0)$
    and $\bm{a}_2=(0,a)$. The yellow-shaded rectangle indicates
    the eight-site unit cell.
    (b) Corresponding first Brillouin zone (BZ) with high-symmetry points
    $\Gamma$, $\mathrm{X}$, $\mathrm{M}$, and $\mathrm{Y}$.
    The reciprocal lattice vectors are $\bm{b}_1=(\frac{\pi}{a}, 0)$
    and $\bm{b}_2=(0, \frac{2\pi}{a})$.
  }\label{fig:2d-ssh-model_DW}
\end{figure}
Figure~\ref{fig:2d-ssh-model_DW}(a) defines the eight-site rectangular unit cell of the multidomain model, and Fig.~\ref{fig:2d-ssh-model_DW}(b) shows the corresponding first Brillouin zone. Each site hosts a single electronic orbital, and the intracell and intercell hoppings retain the amplitudes $-\gamma$ and $-\gamma^{\prime}$, respectively.

The multidomain (MD) Bloch Hamiltonian $\hat{H}_{\rm{MD}}(\mathbf{k})$ is explicitly written as
\begin{widetext}
\begin{equation}
\hat{H}_{\rm{MD}}(\mathbf{k}) =
\begin{pmatrix}
\Delta_{A}       & \rho^{-}_{\gamma}(k_x) & 0  & \rho^{+}_{\gamma^{\prime}}(k_x) & \rho(k_y)  & 0    & 0    & 0\\
\rho^{+}_{\gamma}(k_x)  & \Delta_{B}      & \rho^{-}_{\gamma^{\prime}}(k_x)      & 0 & 0  &\rho(k_y) & 0    & 0\\
0  & \rho^{+}_{\gamma^{\prime}}(k_x)      & \Delta_{C}  & \rho^{-}_{\gamma}(k_x) & 0  & 0            & \rho(k_y)              & 0\\
\rho^{-}_{\gamma^{\prime}}(k_x)           & 0 & \rho^{+}_{\gamma}(k_x)& \Delta_{D} & 0  & 0 & 0      & \rho(k_y)\\
\rho^{\ast}(k_y) & 0    & 0    & 0 & \Delta_{E}               & \rho^{-}_{\gamma}(k_x)    & 0    & \rho^{+}_{\gamma^{\prime}}(k_x)\\
0  & \rho^{\ast}(k_y)   & 0    & 0 & \rho^{+}_{\gamma}(k_x)  & \Delta_{F}  & \rho^{-}_{\gamma^{\prime}}(k_x)  & 0\\
0  & 0   & \rho^{\ast}(k_y)    & 0 & 0  & \rho^{+}_{\gamma^{\prime}}(k_x)  & \Delta_{G}   & \rho^{-}_{\gamma}(k_x)\\
0  & 0   & 0   & \rho^{\ast}(k_y) & \rho^{-}_{\gamma^{\prime}}(k_x)  & 0   & \rho^{+}_{\gamma}(k_x) & \Delta_{H}\\
\end{pmatrix},
\label{eq:HkMD}
\end{equation}
\end{widetext}
where
\begin{align}
& \rho^{\pm}_{\gamma}(k_x)=-\gamma\, e^{\pm i k_{x}\frac{a}{2}},
\qquad
\rho^{\pm}_{\gamma^{\prime}}(k_x)=-\gamma^{\prime} e^{\pm i k_{x}\frac{a}{2}},\\
& \rho(k_y) = -\gamma e^{-i k_y \frac{a}{2}}-\gamma^{\prime}e^{i k_y \frac{a}{2}}.
\end{align}
Here $\rho(k_y)$ coincides with $\rho_y(k_y)$ of Eq.~\eqref{eq:HkSD}, and $\Delta_{\alpha}$ is the on-site potential on sublattice $\alpha$ ($\alpha=A,B,C,D,E,F,G,H$).
Under periodic boundary conditions, the supercell constructed from single
domains $\ctext{1}$ and $\ctext{2}$ contains an
internal head-to-head (H--H) interface crossed by the $BC$ and $FG$ bonds and
a compensating tail-to-tail (T--T) interface at the supercell boundary crossed
by the $AD$ and $EH$ bonds. The supercell constructed from single domains
$\ctext{1}$ and $\ctext{3}$ contains two interfaces
of opposite orientation with no discontinuity in the normal component of the
electronic polarization. These are an internal head-to-tail (H--T) interface and a tail-to-head
(T--H) interface at the supercell boundary. Hereafter, the head-to-head and
head-to-tail arrangements denote the full periodic supercells, whereas H--H,
T--T, H--T, and T--H denote individual interfaces.

\begin{figure}
  \centering
  \includegraphics[width=85mm]{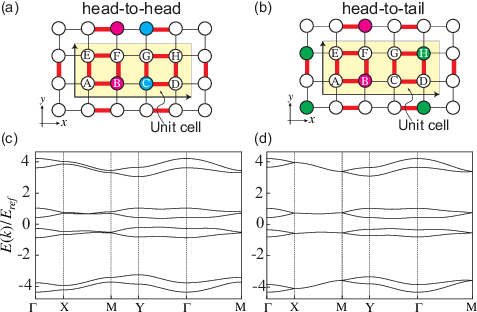}
  \caption{
    (a) Schematic of the periodic head-to-head arrangement, where the on-site potentials are set to
    $\Delta_{A}=\Delta_{D}=\Delta_{E}=\Delta_{F}=\Delta_{G}=\Delta_{H} = \frac{2}{3}$ and
    $\Delta_{B}=\Delta_{C}=-2$.
    (b) Schematic of the periodic head-to-tail arrangement, where the on-site potentials are set to
    $\Delta_{A}=\Delta_{C}=\Delta_{D}=\Delta_{E}=\Delta_{F}=\Delta_{G} = \frac{2}{3}$ and $\Delta_{B}=\Delta_{H}=-2$.
    The hopping parameters are fixed at $\gamma = 1.7$ and $\gamma^{\prime} = 0.3$ in both panels.
    Under periodic boundary conditions, panel (a) contains an internal H--H interface crossed by $BC$ and $FG$ and a compensating T--T interface at the supercell boundary crossed by $AD$ and $EH$.
    Panel (b) contains an internal H--T interface and a T--H interface at the supercell boundary.
    (c), (d) Energy band structures of the head-to-head and head-to-tail arrangements, respectively.
  }\label{fig:2d-ssh-multi-energy_band}
\end{figure}
Figures~\ref{fig:2d-ssh-multi-energy_band}(a) and (b) show the two arrangements, and Figs.~\ref{fig:2d-ssh-multi-energy_band}(c) and (d) their band structures. The head-to-head and head-to-tail periodic arrangements are electronically distinct even though they are constructed from the same polar single-domain building blocks.
In particular, the band structures exhibit a clear difference along the $\mathrm{X}$--$\mathrm{M}$ line. The head-to-head supercell in Fig.~\ref{fig:2d-ssh-multi-energy_band}(c) shows a lifting of the near-degeneracy, whereas the corresponding degeneracy is retained in the head-to-tail supercell of Fig.~\ref{fig:2d-ssh-multi-energy_band}(d).
The two supercells therefore differ in their band dispersions as well as in their symmetries.

\section{Symmetry Constraints on Shift Current}
\label{sec:symmetry-constraints}
The remaining symmetry of the full supercell constrains the allowed components of the third-rank shift-current conductivity tensor $\sigma_{ijk}$ ($i,j,k=x,y$), while the detailed band dispersion and interband optical matrix elements determine their magnitudes and spectral profiles.
This tensor can be expressed in terms of the symmetry operation matrix $\mathcal{R}$ as
\begin{align}
\sigma_{ijk} = \mathcal{R}_{ii'}\mathcal{R}_{jj'}\mathcal{R}_{kk'}\sigma_{i'j'k'}.
\end{align}
The calligraphic symbol $\mathcal{R}$ is used here to distinguish the symmetry
operation matrix from the shift vector $R^{ij}_{nm}$ defined in
Sec.~\ref{sec:single-domain-shift} below.

The head-to-head arrangement of
Fig.~\ref{fig:2d-ssh-multi-energy_band}(a) is invariant under the mirror
operation $\hat{M}_x:\,x\rightarrow \tfrac{3}{2}a-x$, which exchanges the
sublattices as
\begin{align}
A\leftrightarrow D,\quad B\leftrightarrow C,\quad
E\leftrightarrow H,\quad F\leftrightarrow G,
\end{align}
leaving the on-site potentials of
Fig.~\ref{fig:2d-ssh-multi-energy_band}(a) and the alternating
$\gamma/\gamma^{\prime}$ bond pattern unchanged. The matrix representation of
$\hat{M}_x$ is
\begin{align}
\hat{M}_x = \text{diag}(-1,1),
\end{align}
so that components carrying an odd number of $x$ indices vanish.
The following tensor components are symmetry-allowed and may be nonzero:
\begin{align}
\sigma_{xxy}, \sigma_{xyx}, \sigma_{yxx}, \sigma_{yyy}.
\end{align}

By contrast, the head-to-tail arrangement of
Fig.~\ref{fig:2d-ssh-multi-energy_band}(b) is not invariant under any pure
mirror operation, but it is invariant under the nonsymmorphic glide
$\hat{G}_y=\{\hat{M}_y\,|\,\tfrac{1}{2}\bm{a}_1\}$, which combines the mirror
$y\rightarrow \tfrac{1}{2}a-y$ with the fractional translation
$\tfrac{1}{2}\bm{a}_1=(a,0)$. It maps the sublattices as
\begin{align}
\begin{split}
&A\rightarrow G,\quad B\rightarrow H,\quad
 C\rightarrow E,\quad D\rightarrow F,\\
&E\rightarrow C,\quad F\rightarrow D,\quad
 G\rightarrow A,\quad H\rightarrow B,
\end{split}
\end{align}
which leaves the on-site potentials of
Fig.~\ref{fig:2d-ssh-multi-energy_band}(b) invariant. Because $\sigma_{ijk}$
is the macroscopic response in the long-wavelength optical limit
$\mathbf{q}\rightarrow 0$, where $\mathbf{q}$ denotes the response wave vector, the fractional
translation contributes only an overall phase $e^{i\mathbf{q}\cdot
\bm{a}_1/2}\rightarrow 1$ and does not alter the tensor transformation. The
selection rule is therefore governed by the point part
\begin{align}
\hat{M}_y = \text{diag}(1,-1),
\end{align}
so that components carrying an odd number of $y$ indices vanish.
The following tensor components are symmetry-allowed and may be nonzero:
\begin{align}
\sigma_{xyy}, \sigma_{yxy}, \sigma_{yyx}, \sigma_{xxx}.
\end{align}
The fractional translation of the glide does not enter this macroscopic
selection rule, although it remains relevant to the Bloch states at finite
crystal momentum.
Here, we focus on the in-plane components $\sigma_{yyy}$ and $\sigma_{xxx}$. These components are allowed by the symmetries of the head-to-head and head-to-tail arrangements, respectively, and are those excited by illumination polarized along a crystal axis.

The two selection rules are exactly complementary. Since $\sigma_{ijk}$ carries
three Cartesian indices, the numbers of $x$ and $y$ labels satisfy
$n_x+n_y=3$, so precisely one of them is even for every component. A component
allowed in the head-to-head arrangement, which requires $n_x$ even, therefore
has $n_y$ odd and is forbidden in the head-to-tail arrangement, and vice versa.
Each component of $\sigma_{ijk}$ is thus allowed in exactly one of the two
geometries, and the two arrangements admit mutually exclusive sets of tensor
components.
In the long-wavelength limit, this complementarity is unchanged by band dispersion or additional bands that preserve the respective full-supercell symmetries, although it neither fixes the magnitudes of the allowed components nor prevents their cancellation.

This exclusivity applies to the tensor components themselves, not to the
direction of the total photocurrent under arbitrary illumination. In the head-to-head arrangement, for instance, $\sigma_{xxy}$
and $\sigma_{xyx}$ are also symmetry-allowed, so an $x$-directed shift current
$J_x\propto(\sigma_{xxy}+\sigma_{xyx})E_x E_y$ is generated whenever the
incident polarization has both $x$ and $y$ components. This is confirmed
numerically in Appendix~\ref{sec:appendixA} and analyzed as a function of the
polarization angle in Appendix~\ref{sec:appendixB}. The component selection
discussed below therefore refers to illumination polarized along a crystal
axis.

\section{Shift Current in Single-Domain Structures}
\label{sec:single-domain-shift}
We first summarize the shift-current formalism and then calculate the responses of the three single-domain configurations in Fig.~\ref{fig:single-domain-polarization}.
In Sec.~\ref{sec:multidomain-shift}, we compare each multidomain response with the equal-area average of its two constituent single-domain responses.

The shift-current sheet density $J^{\rm{shift}}_i$, with units of $\mathrm{A/m}$, is related to the Fourier components $E_j(\omega)$ and $E_k(-\omega)$ of the incident electric field, where $\omega$ is the angular frequency of the light, through the 2D sheet-conductivity tensor $\sigma_{ijk}$ as follows:
\begin{equation}
J^{\rm{shift}}_i=2\sum_{jk}\sigma_{ijk}(0;\,\omega,-\omega)E_j(\omega)E_k(-\omega),
\end{equation}
Here a tilde denotes a time-dependent field, so that the real-time electric-field component is
$\tilde{E}_{j(k)}(t)=E_{j(k)}(\omega)e^{-i\omega t}+E_{j(k)}(-\omega)e^{i\omega t}$.
We calculate the shift-current conductivity $\sigma_{ijk}$ using quadratic response theory~\cite{SipeGhahramani1993,AversaSipe1995,HughesSipe1996,SipeShkrebtii2000,NastosSipe2010,IbanezAzpiroz2018}.
\begin{widetext}
\begin{equation}
  \sigma_{ijk}(0;\,\omega,-\omega)
  =-\frac{i\hbar e^{3}}{2S}
   \sum_{\mathbf{k}}\sum_{m n}
   \frac{ f_{nm}\!\left[
      \mathcal{I}_{mn}^{jk}(\mathbf{k})\,R_{nm}^{ik}(\mathbf{k})
      + \mathcal{I}_{mn}^{kj}(\mathbf{k})\,R_{nm}^{ij}(\mathbf{k})
   \right]}
   { E_{mn}^{2}\,\bigl(E_{mn}-\hbar\omega-i\eta\bigr)},
\end{equation}
\end{widetext}
Here $i$ denotes the direction of the shift current, and $j$ and $k$ denote the directions of the incident electric field.
Throughout, the italic subscripts $i$, $j$, and $k$ label Cartesian directions, whereas the boldface $\mathbf{k}$ denotes the crystal momentum.
Electric-field components are written with a Cartesian subscript and a frequency argument, as in $E_j(\omega)$, which distinguishes them from the band energies $E_n(\mathbf{k})$.
The quantity $S$ is the area of the system, $e$ is the elementary charge, and $\eta$ is a phenomenological broadening that regularizes the resonant denominator.
The indices $n$ and $m$ label bands, and $f$ is the Fermi--Dirac distribution
function. We define
$f_{nm}\equiv f\!\bigl(E_n(\mathbf{k})\bigr)-f\!\bigl(E_m(\mathbf{k})\bigr)$
and, more generally, $E_{ab}\equiv E_a(\mathbf{k})-E_b(\mathbf{k})$ for any band indices $a$ and $b$.

All calculations are performed for the spinless model at half
filling and zero temperature, $T=0$, with the chemical potential fixed in the
central gap at $\mu/E_{\rm ref}=0$. Accordingly,
$f(E)=\Theta(\mu-E)$, where $\Theta$ is the Heaviside step function. The two lower bands of each four-band single-domain
Hamiltonian and the four lower bands of each eight-band multidomain
Hamiltonian are occupied. We use the broadening $\eta/E_{\rm ref}=0.01$
throughout.

We now define the two ingredients of the conductivity, the transition intensity and the shift vector.
The transition intensity $\mathcal{I}^{jk}_{mn}$ is defined as
\begin{equation}
  \mathcal{I}^{jk}_{mn}(\mathbf{k}) = v^{\,j}_{mn}(\mathbf{k})\,v^{\,k}_{nm}(\mathbf{k}),
\end{equation}
where $v^{\,j}_{mn}=\frac{1}{\hbar}\langle m \lvert \partial_{k_j}\hat H\rvert n\rangle$ denotes the velocity matrix element.
Here, $|n\rangle$ and $|m\rangle$ are shorthand for the Bloch states $\lvert u_n(\mathbf{k})\rangle$ and $\lvert u_m(\mathbf{k})\rangle$, respectively.
$R^{ij}_{nm}$ is the shift vector, given by
\begin{widetext}
\begin{equation}
\begin{split}
  R^{ij}_{nm}(\mathbf{k})
  = \frac{i\hbar}{v^{\,j}_{nm}}
  \Bigg(
      \frac{v^{\,i}_{nm}\bigl(v^{\,j}_{mm}-v^{\,j}_{nn}\bigr)}{E_{nm}}
    + \frac{v^{\,j}_{nm}\bigl(v^{\,i}_{mm}-v^{\,i}_{nn}\bigr)}{E_{nm}}
    + \frac{w^{\,ji}_{nm}}{\hbar}
    + \sum_{l\neq n,m}\Bigl(
        \frac{v^{\,i}_{nl}v^{\,j}_{lm}}{E_{nl}}
        - \frac{v^{\,j}_{nl}v^{\,i}_{lm}}{E_{lm}}
      \Bigr)
  \Bigg),
\end{split}
\end{equation}
\end{widetext}
with $w^{\,ji}_{nm}=\frac{1}{\hbar}\langle n \lvert \partial_{k_j}\partial_{k_i}\hat H\rvert m\rangle$~\cite{AversaSipe1995,SipeShkrebtii2000,IbanezAzpiroz2018}.
Separating the response into the transition intensity and the shift vector in this way also allows the interband transitions that dominate a given spectral feature to be identified~\cite{Kusunoki2026SnSTB}.

The intermediate-state sum in $R^{ij}_{nm}$ runs over the band index $l$ and
excludes $l=n$ and $l=m$, which
would otherwise produce vanishing energy denominators $E_{nn}=E_{mm}=0$.
Although $R^{ij}_{nm}$ carries an explicit factor $1/v^{\,j}_{nm}$ and is
therefore singular wherever $v^{\,j}_{nm}$ vanishes, this factor cancels
identically against the transition intensity in the combination that enters
the conductivity,
\begin{equation}
  \mathcal{I}^{jk}_{mn}R^{ik}_{nm}
  =v^{\,j}_{mn}v^{\,k}_{nm}\,
   \frac{i\hbar}{v^{\,k}_{nm}}\bigl(\cdots\bigr)
  =i\hbar\,v^{\,j}_{mn}\bigl(\cdots\bigr),
  \label{eq:regularity}
\end{equation}
where $(\cdots)$ denotes the bracket of $R^{ik}_{nm}$. The integrand of
$\sigma_{ijk}$ is thus regular at $v^{\,j}_{nm}=0$, and no numerical
regularization of the shift vector is required. The same cancellation holds
for the bond-resolved decomposition of Sec.~\ref{sec:real-space}, in which the
prefactor $1/v^{\,j}_{nm}$ retains the total velocity matrix element.

For incident light linearly polarized along the Cartesian axis $j$, the shift current is governed by the components
\begin{equation}
  \sigma_{ijj}(0;\,\omega,-\omega)
  =-\frac{i\hbar e^{3}}{S}
   \sum_{\mathbf{k}}\sum_{m n}
   \frac{ f_{nm}\,\mathcal{I}_{mn}^{jj}(\mathbf{k})\,R_{nm}^{ij}(\mathbf{k})}
   { E_{mn}^{2}\,\bigl(E_{mn}-\hbar\omega-i\eta\bigr)}.
   \label{shift_current_conductivity_ijj}
\end{equation}
The Brillouin-zone sum is normalized by the system area $S$, and in the
thermodynamic limit
\begin{equation}
  \frac{1}{S}\sum_{\mathbf{k}}\;\longrightarrow\;
  \int_{\rm BZ}\frac{d^2k}{(2\pi)^2},
  \label{eq:ksum-continuum}
\end{equation}
so that $\sigma_{ijk}$ is an intensive 2D sheet conductivity, independent of
the number of supercells. Because the single-domain and multidomain
Brillouin zones differ in size, this normalization ensures that the
single-domain and multidomain spectra are directly comparable. Numerically, we
evaluate the Brillouin-zone integrals on uniform $300\times300$
$\mathbf{k}$-point meshes for the four-band single-domain Hamiltonians and on
uniform $200\times200$ and $1000\times1000$ meshes for the eight-band
head-to-head and head-to-tail Hamiltonians, respectively. The single-domain
and head-to-head spectra are computed at 500 uniformly spaced photon energies
over $0\leq\hbar\omega/E_{\rm ref}\leq9$, corresponding to a spacing
$\Delta(\hbar\omega/E_{\rm ref})=9/499\simeq0.0180$, and the head-to-tail
spectra at 250 uniformly spaced photon energies over the same range,
corresponding to
$\Delta(\hbar\omega/E_{\rm ref})=9/249\simeq0.0361$. For
$\eta/E_{\rm ref}=0.01$, the spectra are converged with respect to the
$\mathbf{k}$-point density.

We report the conductivity in units of the 2D sheet-conductivity scale
\begin{equation}
  \sigma_{0}^{\rm 2D}=\frac{e^3 a}{2\hbar E_{\rm ref}},
  \label{eq:2d-conductivity-unit}
\end{equation}
which has the physical dimensions $\mathrm{A\,m\,V^{-2}}$, as required for a
sheet current density $J^{\rm shift}_i$ measured in $\mathrm{A/m}$ and driven
by an in-plane field measured in $\mathrm{V/m}$. The numerical values shown
below are therefore the dimensionless ratios $\sigma_{ijk}/\sigma_0^{\rm 2D}$, and setting
$a=E_{\rm ref}=1$ does not alter these values.

We now examine the shift-current spectra of the three representative single-domain configurations defined in Fig.~\ref{fig:single-domain-polarization}.
\begin{figure*}
  \centering
  \includegraphics[width=170mm]{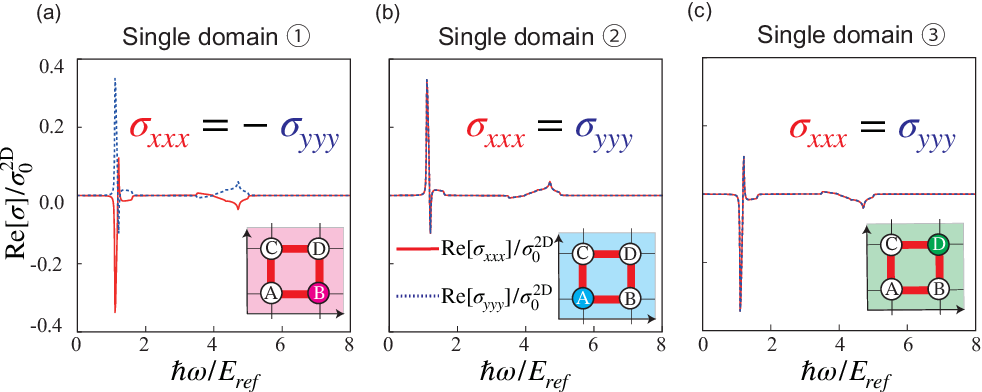}
  \caption{
    Real parts of $\sigma_{xxx}$ (red solid lines) and $\sigma_{yyy}$ (blue dotted lines) for (a) single domain $\ctext{1}$, (b) single domain $\ctext{2}$, and (c) single domain $\ctext{3}$.
    The insets show the corresponding unit-cell configurations.
    The conductivity is reported as $\mathrm{Re}[\sigma_{ijk}]/\sigma_0^{\rm 2D}$.
  }
  \label{shift_current_single_domain}
  \end{figure*}
Figure~\ref{shift_current_single_domain} shows how the on-site potential pattern of a single domain controls the relation between $\sigma_{xxx}$ and $\sigma_{yyy}$.
Panel (a) shows that $\mathrm{Re}[\sigma_{xxx}] = -\mathrm{Re}[\sigma_{yyy}]$ for single domain $\ctext{1}$. By contrast, panels (b) and (c) show that $\mathrm{Re}[\sigma_{xxx}] = \mathrm{Re}[\sigma_{yyy}]$ for single domains $\ctext{2}$ and $\ctext{3}$, respectively.

These relations are not accidental. Each follows from a mirror symmetry through
one of the diagonals of the corresponding unit cell. Single domains $\ctext{2}$ and
$\ctext{3}$, whose deep on-site potential sits on site $A$ and on site $D$,
respectively, are both invariant under the mirror
$\hat{M}_{d}:(x,y)\rightarrow(y,x)$ through the cell diagonal, which exchanges
$B\leftrightarrow C$ and leaves $A$ and $D$ fixed. Its matrix representation is
$\mathcal{R}=\left(\begin{smallmatrix}0&1\\1&0\end{smallmatrix}\right)$, which interchanges the
Cartesian labels, so that
\begin{align}
\sigma_{xxx}=\sigma_{yyy}
\qquad(\text{single domains }\ctext{2}\text{ and }\ctext{3}).
\end{align}
Single domain $\ctext{1}$, whose deep site is $B$, is instead invariant under
the antidiagonal mirror $\hat{M}_{\bar{d}}:(x,y)\rightarrow(\tfrac{1}{2}a-y,\,
\tfrac{1}{2}a-x)$, which exchanges $A\leftrightarrow D$ and leaves $B$ and $C$
fixed. The corresponding matrix representation is
$\mathcal{R}=\left(\begin{smallmatrix}0&-1\\-1&0\end{smallmatrix}\right)$, and the three
sign factors carried by a third-rank tensor give
\begin{align}
\sigma_{xxx}=-\sigma_{yyy}
\qquad(\text{single domain }\ctext{1}).
\end{align}
Single domains $\ctext{2}$ and $\ctext{3}$ are furthermore related to each
other by spatial inversion about the point $(\tfrac{1}{4}a,\tfrac{1}{4}a)$,
which maps $A\leftrightarrow D$ and $B\leftrightarrow C$ and carries the
on-site pattern of $\ctext{2}$ into that of $\ctext{3}$. Since $\sigma_{ijk}$
is odd under inversion, this implies
$\sigma^{\ctext{3}}_{ijk}=-\sigma^{\ctext{2}}_{ijk}$, consistent with the
opposite electronic polarizations quoted in
Fig.~\ref{fig:single-domain-polarization} and with the mutually reversed
spectra of Figs.~\ref{shift_current_single_domain}(b) and (c).
The three single-domain structures therefore differ not only in the
distribution of spectral weight between the $x$ and $y$ directions, but also
in the overall sign of the response. Their spectral line shapes are related by
the exact symmetry operations above rather than being independent.
The peaks around $\hbar\omega/E_{\rm ref}=1.1$ are attributed to interband transitions across the smallest gap, that is, between the second and third bands.
The sharpness of these peaks depends on the chosen model parameters and broadening and is not a quantitative prediction for monolayer GeS or SnS.

\section{Shift Current in Multidomain Structures}
\label{sec:multidomain-shift}
We consider two representative periodic interface arrangements assembled from the single domains described above.
These are the head-to-head and head-to-tail arrangements shown in Figs.~\ref{shift_current_multi_domain_headtohead}(a) and~\ref{shift_current_multi_domain_headtotail}(a), respectively.
On the common polarization branch adopted here, a discontinuity in the normal
component of the electronic polarization defines a formal bound charge
inferred from the electronic polarization
texture~\cite{VanderbiltKingSmith1993}. In the head-to-head
arrangement, the internal H--H interface carries a positive formal bound
charge, accompanied by an equal and opposite one at the compensating T--T
interface at the periodic boundary, so that the supercell remains charge
neutral overall. In the head-to-tail arrangement, the normal components of the
electronic polarization are identical across both oppositely oriented H--T and
T--H interfaces, so the discontinuity vanishes identically in the present
model. These bound charges are formal electronic quantities. The associated
electrostatic response of a real interface is outside the model's scope,
as specified in Sec.~\ref{sec:model-polarization}.

\subsection{Head-to-head periodic arrangement}
\begin{figure}
  \centering
  \includegraphics[width=85mm]{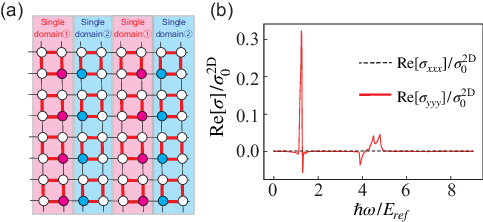}
  \caption{
    (a) Schematic of the periodic head-to-head arrangement, in which H--H and T--T interfaces alternate under periodic boundary conditions.
    (b) Real parts of $\sigma_{xxx}$ (black dotted line) and $\sigma_{yyy}$ (red solid line).
    The conductivity is reported as $\mathrm{Re}[\sigma_{ijk}]/\sigma_0^{\rm 2D}$.
  }
  \label{shift_current_multi_domain_headtohead}
  \end{figure}
In this arrangement the polarization vectors of neighboring domains point toward the internal H--H interface.
Figure~\ref{shift_current_multi_domain_headtohead}(b) shows that
$\mathrm{Re}[\sigma_{xxx}]$, which is the response for fields polarized perpendicular to the interfaces, vanishes,
whereas $\mathrm{Re}[\sigma_{yyy}]$, which is the response for fields polarized parallel to the interfaces, remains finite.
The macroscopic (area-averaged) polarization normal to the interfaces cancels
because the two equally sized domains carry opposite normal components.

\subsection{Head-to-tail periodic arrangement}
\begin{figure}
  \centering
  \includegraphics[width=85mm]{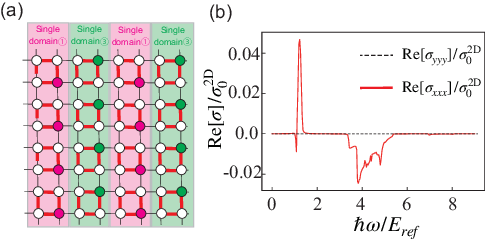}
  \caption{
    (a) Schematic of the periodic head-to-tail arrangement, in which oppositely oriented H--T and T--H interfaces alternate under periodic boundary conditions.
    (b) Real parts of $\sigma_{xxx}$ (red solid line) and $\sigma_{yyy}$ (black dotted line).
    The conductivity is reported as $\mathrm{Re}[\sigma_{ijk}]/\sigma_0^{\rm 2D}$.
  }
  \label{shift_current_multi_domain_headtotail}
  \end{figure}
In this arrangement the two domains meet without pointing their polarizations directly toward each other.
At either interface in Fig.~\ref{shift_current_multi_domain_headtotail}(a), the discontinuity in the
normal component of the electronic polarization vanishes identically in the
present model. Because this arrangement has a different full-supercell
symmetry from the head-to-head case, as established in
Sec.~\ref{sec:symmetry-constraints}, it is expected to select a different set
of shift-current tensor components.

Figure~\ref{shift_current_multi_domain_headtotail}(b) shows the complementary behavior.
Here $\mathrm{Re}[\sigma_{yyy}]$, the response for parallel polarization, vanishes,
whereas $\mathrm{Re}[\sigma_{xxx}]$, the response for perpendicular polarization, remains finite.
The macroscopic (area-averaged) polarization component parallel to the interfaces cancels, whereas the normal component remains finite.

For the supercells considered here, the maximum value of $\mathrm{Re}[\sigma_{xxx}]/\sigma_0^{\rm 2D}$ for the head-to-tail arrangement is approximately
$0.05$, roughly seven times smaller than the single-domain maximum ($\simeq0.34$).
The head-to-head arrangement, by contrast, retains a magnitude comparable to the single-domain value [Fig.~\ref{shift_current_multi_domain_headtohead}(b)].
At this width, the two arrangements therefore differ not only in which response direction is selected, but also in the magnitude of the surviving component. These peak magnitudes are specific to the domain width considered here. The spectra for wider domains are presented in Appendix~\ref{sec:appendixC}.
Accordingly, the head-to-head and head-to-tail arrangements should be viewed as symmetry-distinct periodic supercells.
The former selects the response parallel to the interfaces, whereas the latter selects the response perpendicular to the interfaces.
Whether an equal-area independent-domain average reproduces the detailed spectrum is tested below.

\subsection{Comparison with area-averaged single-domain responses}
To test the independent-domain approximation for the supercells considered here, we compare each multidomain response with the equal-area average of the corresponding constituent single-domain responses. For the two equally sized domains in the present supercell, we define
\begin{align}
\sigma_{yyy,\mathrm{avg}}
&=
\frac{\sigma_{yyy}^{\ctext{1}}+\sigma_{yyy}^{\ctext{2}}}{2},
\\
\sigma_{xxx,\mathrm{avg}}
&=
\frac{\sigma_{xxx}^{\ctext{1}}+\sigma_{xxx}^{\ctext{3}}}{2},
\end{align}
where the circled superscripts denote the single-domain configurations shown
in Fig.~\ref{fig:single-domain-polarization}. The first and second expressions
correspond to the head-to-head and head-to-tail arrangements, respectively.

\begin{figure*}
  \centering
  \includegraphics[width=170mm]{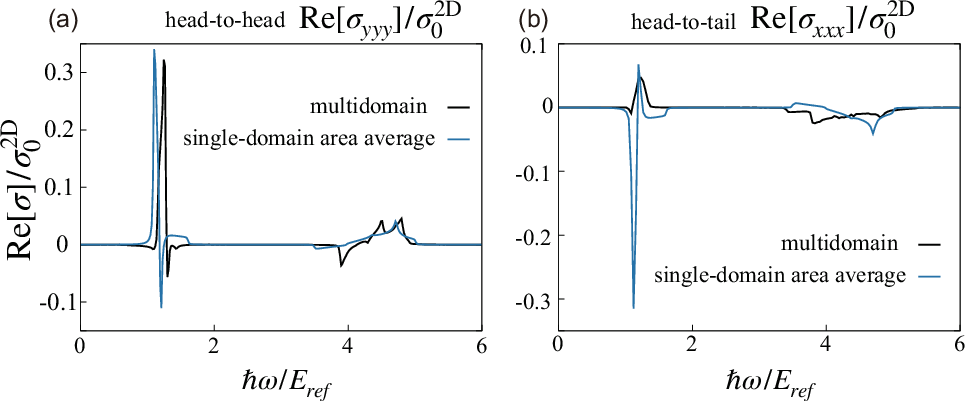}
  \caption{
  Direct comparison of the multidomain responses for the supercells studied in the main text (black solid lines) with equal-area averages of the constituent single-domain responses (blue solid lines):
  (a) $\mathrm{Re}[\sigma_{yyy}]$ for the head-to-head arrangement and
  (b) $\mathrm{Re}[\sigma_{xxx}]$ for the head-to-tail arrangement.
  The conductivity is reported as $\mathrm{Re}[\sigma_{ijk}]/\sigma_0^{\rm 2D}$.
  }
  \label{fig:area_average}
\end{figure*}

Figure~\ref{fig:area_average} compares each multidomain spectrum with the corresponding equal-area average. For the head-to-head
arrangement [Fig.~\ref{fig:area_average}(a)], the equal-area average
reproduces the sign and the approximate height of the leading peak, but not
its position, the adjacent sign-changing structure, or the additional features
near $\hbar\omega/E_{\rm ref}\simeq 4$. For the head-to-tail arrangement
[Fig.~\ref{fig:area_average}(b)], the disagreement is qualitative rather than
quantitative. For the supercell considered here, at the leading spectral feature the equal-area average has the
opposite sign to the multidomain result and a magnitude several times larger.
Thus, for the domain width considered here, the detailed response of either periodic multidomain arrangement is not
reproduced by this equal-area independent-domain average, and the head-to-tail
case demonstrates that even the sign of the dominant feature cannot be
inferred from the constituent single-domain spectra.

This comparison tests the independent-domain approximation for the supercells
studied in the main text, where each constituent domain is one primitive cell wide and
the two interfaces are strongly coupled. It does not isolate a domain-wall
contribution.

In Appendix~\ref{sec:appendixA}, we present additional tensor components of the shift-current response for the head-to-head arrangement and compare them with the symmetry constraints.
In Appendix~\ref{sec:appendixB}, we analyze the incident-light polarization dependence of the shift-current conductivity.
In Appendix~\ref{sec:appendixC}, we examine the width dependence of
the multidomain spectra themselves and their bond-resolved spatial patterns.
An isolated-wall response is not extracted from either comparison.

\section{Bond-Resolved Interpretation of the Shift Current}
\label{sec:real-space}
We introduce a bond-resolved decomposition that assigns portions of the output-current vertex to hopping channels between localized orbitals, thereby providing a real-space perspective on how different bonds combine in the multidomain response. This decomposition is an interpretive,
basis-dependent partition of the total response.
It depends on the choice of
localized-orbital basis, the unit-cell convention, and the partition of the
Hamiltonian into hopping channels. Only the summed conductivity is a directly
observable quantity, and the individual channel contributions should not be
read as locally measurable currents.

\begin{table}[t]
\centering
\caption{
Hopping channels used in the bond-resolved shift-current
analysis. Although the intracell and intercell channels along
the \(y\) direction connect the same pairs of site orbitals,
they correspond to different lattice displacements and are
therefore treated as distinct hopping channels.
}
\label{tab:hopping_channels}
\begin{ruledtabular}
\begin{tabular}{c c c}
Direction & Intracell channels & Intercell channels \\
\hline
\(x\)
&
\(AB,\;CD,\;EF,\;GH\)
&
\(BC,\;FG,\;AD,\;EH\)
\\
\(y\)
&
\(AE,\;BF,\;CG,\;DH\)
&
\(AE,\;BF,\;CG,\;DH\)
\end{tabular}
\end{ruledtabular}
\end{table}

We introduce a hopping-channel index
\(\lambda\), which specifies the hopping direction, the pair
of connected sites, and the intracell or intercell character
of the hopping. The set of hopping channels is denoted by
\(\Lambda\), with \(|\Lambda|=16\) in the present
eight-site model. The sixteen channels are summarized in
Table~\ref{tab:hopping_channels}.
The Bloch Hamiltonian is decomposed as
\begin{align}
\hat{H}(\mathbf{k})
=
\hat{H}_{\text{on-site}}
+
\sum_{\lambda\in\Lambda}
\hat{H}_{\lambda}(\mathbf{k}).
\label{eq:H_channel_decomposition}
\end{align}
Each hopping channel \(\lambda\) specifies two endpoint orbitals
\(\alpha(\lambda),\beta(\lambda)
\in\{A,B,C,D,E,F,G,H\}\), a hopping amplitude
\(t_{\lambda}\), and a real-space displacement
\(\boldsymbol{\delta}_{\lambda}\).
The corresponding contribution to the Bloch Hamiltonian is
written as
\begin{align}
\hat{H}_{\lambda}(\mathbf{k})
=
t_{\lambda}
e^{i\mathbf{k}\cdot\boldsymbol{\delta}_{\lambda}}
|\alpha(\lambda)\rangle
\langle\beta(\lambda)|
+
\mathrm{H.c.},
\label{eq:H_channel}
\end{align}
where \(\boldsymbol{\delta}_{\lambda}\) is the displacement
associated with the hopping from \(\beta(\lambda)\) to
\(\alpha(\lambda)\), including the unit-cell translation for
an intercell hopping. Channels connecting the same pair of
site orbitals are distinguished by their different
\(\boldsymbol{\delta}_{\lambda}\), and hence by their
intracell or intercell character.

Because the on-site term $\hat{H}_{\text{on-site}}$ is independent of momentum, the
Hamiltonian derivatives satisfy
\begin{align}
\partial_{k_i}\hat{H}(\mathbf{k})
&=
\sum_{\lambda\in\Lambda}
\partial_{k_i}\hat{H}_{\lambda}(\mathbf{k}),
\\
\partial_{k_j}\partial_{k_i}\hat{H}(\mathbf{k})
&=
\sum_{\lambda\in\Lambda}
\partial_{k_j}\partial_{k_i}
\hat{H}_{\lambda}(\mathbf{k}).
\end{align}
We define the hopping-channel-resolved velocity and
generalized velocity matrix elements as
\begin{align}
v^i_{mn;\lambda}(\mathbf{k})
&=
\frac{1}{\hbar}
\langle m|
\partial_{k_i}\hat{H}_{\lambda}(\mathbf{k})
|n\rangle,
\label{eq:v_channel}
\\
w^{ji}_{mn;\lambda}(\mathbf{k})
&=
\frac{1}{\hbar}
\langle m|
\partial_{k_j}\partial_{k_i}
\hat{H}_{\lambda}(\mathbf{k})
|n\rangle.
\label{eq:w_channel}
\end{align}

The corresponding total matrix elements are recovered as
\begin{align}
v^i_{mn}(\mathbf{k})
&=
\sum_{\lambda\in\Lambda}
v^i_{mn;\lambda}(\mathbf{k}),
\\
w^{ji}_{mn}(\mathbf{k})
&=
\sum_{\lambda\in\Lambda}
w^{ji}_{mn;\lambda}(\mathbf{k}).
\label{eq:vw_channel_sum}
\end{align}
The shift-current conductivity is consequently decomposed as
\begin{align}
\sigma_{ijj}(0;\omega,-\omega)
=
\sum_{\lambda\in\Lambda}
\sigma_{ijj;\lambda}(0;\omega,-\omega),
\label{eq:sigma_channel_sum}
\end{align}
where
\begin{align}
\sigma_{ijj;\lambda}(0;\omega,-\omega)
=
-\frac{i\hbar e^3}{S}
\sum_{\mathbf{k}}\sum_{mn}
\frac{
f_{nm}\,
\mathcal{I}^{jj}_{mn}(\mathbf{k})\,
R^{ij}_{nm;\lambda}(\mathbf{k})
}{
E_{mn}^{2}
\left(
E_{mn}-\hbar\omega-i\eta
\right)
}.
\label{eq:shift_local}
\end{align}
Here,
\begin{widetext}
\begin{align}
  R_{nm;\lambda}^{ij}(\mathbf{k})
  = \frac{i\hbar}{v^{j}_{nm}}
  \Bigg(
      \frac{v^{i}_{nm;\lambda}\bigl(v^{j}_{mm}-v^{j}_{nn}\bigr)}{E_{nm}}
    + \frac{v^{j}_{nm}\bigl(v^{i}_{mm;\lambda}-v^{i}_{nn;\lambda}\bigr)}{E_{nm}}
    + \frac{w^{ji}_{nm;\lambda}}{\hbar}
    + \sum_{l\neq n,m}\Bigl(
        \frac{v^{i}_{nl;\lambda}v^{j}_{lm}}{E_{nl}}
        - \frac{v^{j}_{nl}v^{i}_{lm;\lambda}}{E_{lm}}
      \Bigr)
  \Bigg).
\end{align}
\end{widetext}
The optical vertices associated with the incident-field direction \(j\)
are kept as total matrix elements, whereas the
output-current vertex along the \(i\) direction is resolved
into hopping-channel contributions.
As emphasized above, each \(\sigma_{ijj;\lambda}\) should be interpreted as a
bookkeeping contribution to the total conductivity, rather than as an
independently measurable local observable.

\begin{figure*}
  \centering
  \includegraphics[width=167mm]{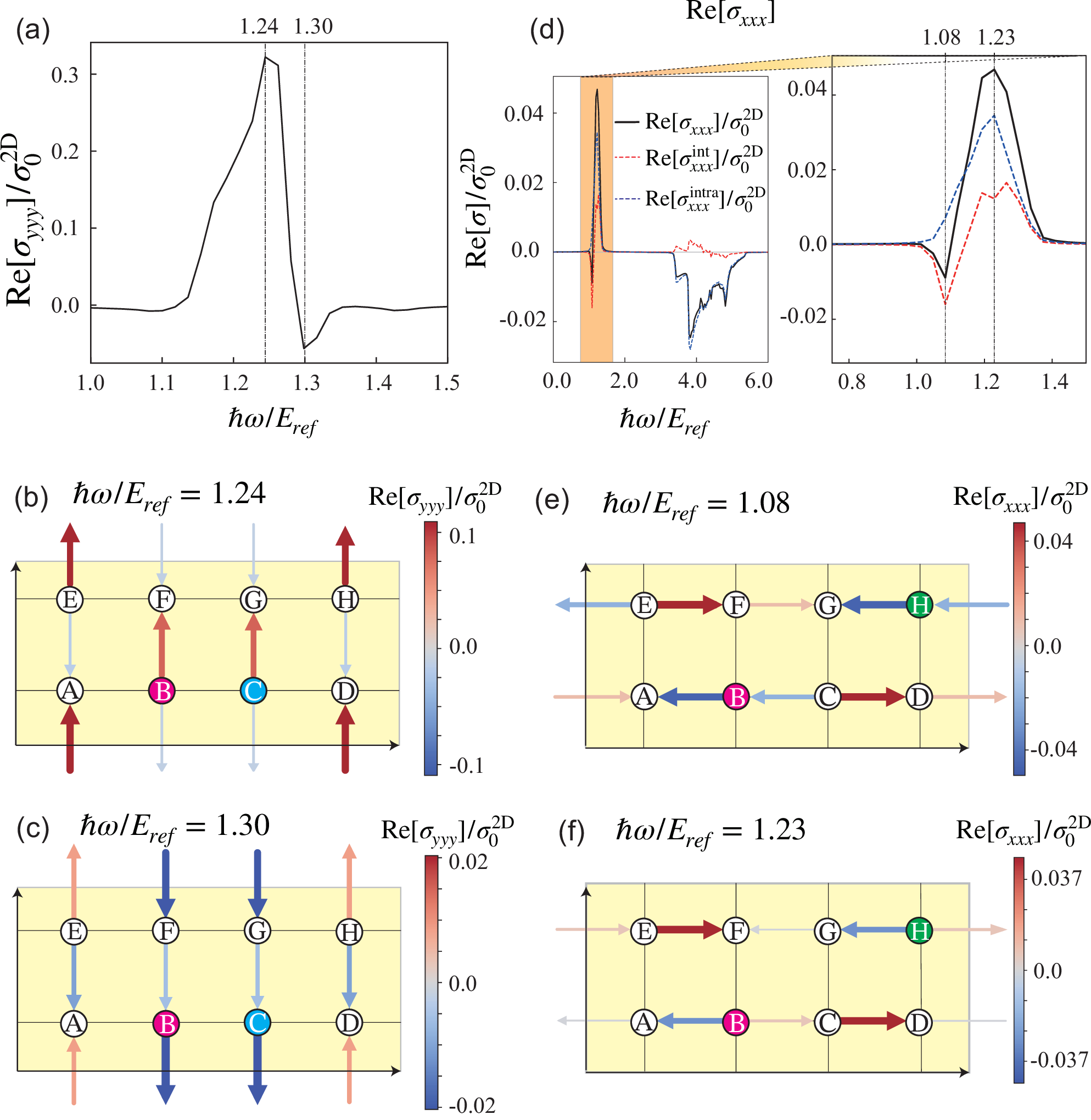}
  \caption{
(a) Enlarged view of the real part of $\sigma_{yyy}$ for the head-to-head arrangement in the photon-energy range $1.0\leq\hbar\omega/E_{\rm ref}\leq1.5$.
(b), (c) Schematic real-space representations of the bond-resolved contributions at
(b) $\hbar\omega/E_{\rm ref}=1.24$ and (c) $\hbar\omega/E_{\rm ref}=1.30$, respectively.
(d) Real part of $\sigma_{xxx}$ (black solid line), the interface-crossing-bond contribution $\sigma_{xxx}^{\rm int}$ (red dashed line), and the intradomain-bond contribution $\sigma_{xxx}^{\rm intra}$ (blue dashed line) for the head-to-tail arrangement, with the selected photon-energy range enlarged on the right.
(e), (f) Schematic real-space representations of the bond-resolved contributions at
(e) $\hbar\omega/E_{\rm ref}=1.08$ and (f) $\hbar\omega/E_{\rm ref}=1.23$, respectively.
The schematics cover the full periodic supercell. In panels (b) and (c), the internal H--H interface lies between $B$ and $C$ and between $F$ and $G$, while the compensating T--T interface lies between $D$ and $A$ and between $H$ and $E$ across the periodic boundary. The corresponding interfaces in panels (e) and (f) are the oppositely oriented H--T and T--H interfaces.
In panels (b), (c), (e), and (f), each arrow is drawn on the bond of the
corresponding hopping channel $\lambda$. The arrow points along $+y$ ($+x$)
for a positive bond-resolved contribution $\sigma_{yyy;\lambda}$
($\sigma_{xxx;\lambda}$) and along $-y$ ($-x$) for a negative one. The arrow
length is fixed by the corresponding bond geometry and does not encode the
magnitude of the contribution. The arrow width represents its magnitude,
while the arrow color gives its signed value according to the accompanying
color scale. The arrow widths and color scales are normalized independently
within each panel. Quantitative comparisons between panels should therefore
be made using the numerical color-bar ranges.
The conductivity and bond-resolved contributions are reported as
$\mathrm{Re}[\sigma]/\sigma_0^{\rm 2D}$.
  }
  \label{fig:localized_current}
\end{figure*}
Figure~\ref{fig:localized_current}(a) shows an enlarged view of $\mathrm{Re}[\sigma_{yyy}]$ for the head-to-head arrangement, and Figs.~\ref{fig:localized_current}(b) and (c) the corresponding bond-resolved patterns at the two photon energies marked there.
At $\hbar\omega/E_{\rm ref}=1.24$ [Fig.~\ref{fig:localized_current}(b)], where $\mathrm{Re}[\sigma_{yyy}]$ takes a large positive value, the positive bond-resolved contributions on the vertical bonds dominate over the negative ones.
Their incomplete cancellation produces a large positive macroscopic response.
By contrast, at $\hbar\omega/E_{\rm ref}=1.30$ [Fig.~\ref{fig:localized_current}(c)], $\mathrm{Re}[\sigma_{yyy}]$ changes sign.
The bond-resolved pattern also changes, with negative contributions becoming dominant and reversing the sign of $\mathrm{Re}[\sigma_{yyy}]$ across the spectral peak.
Only the vertical bonds appear in Figs.~\ref{fig:localized_current}(b) and
(c). This is exact rather than a selection made for clarity.
From Eq.~\eqref{eq:H_channel}, the channel-resolved output vertex
$\partial_{k_y}\hat{H}_{\lambda}$ is proportional to the $y$ component of the
displacement $\bm{\delta}_{\lambda}$, which vanishes for every $x$-oriented
hopping channel. Consequently, the $y$-directed output current receives
contributions only from $y$-oriented channels, none of which crosses an
interface. The horizontal bonds still influence the response, but
through the optical vertices and the eigenstates rather than through the
output-current vertex.
The head-to-head periodic supercell nonetheless contains bonds that cross both
interfaces. The $BC$ and $FG$ bonds cross the internal H--H interface, while $AD$ and
$EH$ cross the compensating T--T interface through the periodic boundary.
Because the two interfaces in the supercell considered here are separated by only
one primitive cell, we present the individual vertical-bond pattern rather
than assigning it to a single isolated interface. For the wider domains in
Appendix~\ref{sec:appendixC}, we also group these vertical-bond contributions
by their proximity to the two interfaces.

Figure~\ref{fig:localized_current}(d) shows $\mathrm{Re}[\sigma_{xxx}]$ for the head-to-tail arrangement, decomposed into an interface-crossing-bond contribution ($\sigma_{xxx}^{\rm int}$) and an intradomain-bond contribution ($\sigma_{xxx}^{\rm intra}$).
The interface-crossing contribution is the sum over bonds crossing either interface of the periodic supercell, and the intradomain contribution is the sum over bonds inside either constituent domain, so that
\begin{equation}
\sigma_{xxx} = \sigma_{xxx}^{\rm int} + \sigma_{xxx}^{\rm intra},
\end{equation}
with
\begin{align}
\sigma_{xxx}^{\rm int} &= \sigma_{xxx;BC}+\sigma_{xxx;FG}+\sigma_{xxx;AD}+\sigma_{xxx;EH},\\
\sigma_{xxx}^{\rm intra} &= \sigma_{xxx;AB}+\sigma_{xxx;CD}+\sigma_{xxx;EF}+\sigma_{xxx;GH}.
\end{align}
Here the bond labels denote the corresponding \(x\)-direction hopping channels \(\lambda\) listed in Table~\ref{tab:hopping_channels}, so that, for example, \(\sigma_{xxx;BC}\) is \(\sigma_{xxx;\lambda}\) for the \(BC\) channel.
In Fig.~\ref{fig:localized_current}(d), the black solid, red dashed, and blue dashed lines represent $\sigma_{xxx}$, $\sigma_{xxx}^{\rm int}$, and $\sigma_{xxx}^{\rm intra}$, respectively.
Figures~\ref{fig:localized_current}(e) and (f) show the corresponding real-space bond-resolved patterns at the two photon energies marked in Fig.~\ref{fig:localized_current}(d).
For the supercell considered here, at $\hbar\omega/E_{\rm ref}=1.08$ [Fig.~\ref{fig:localized_current}(e)], the interface-crossing-bond contribution is negative, whereas the intradomain-bond contribution is positive.
Their partial cancellation reduces the total $\sigma_{xxx}$. This sign combination is specific to the width and frequency shown. The wider-domain patterns are discussed in Appendix~\ref{sec:appendixC}.
By contrast, at $\hbar\omega/E_{\rm ref}=1.23$ [Fig.~\ref{fig:localized_current}(f)], both contributions become positive, and the intradomain-bond contribution dominates the total response, producing the spectral peak in $\mathrm{Re}[\sigma_{xxx}]$.
Thus, the constructive or destructive combination of the intradomain-bond and interface-crossing-bond contributions determines the spectral profile of the finite $\sigma_{xxx}$ response, providing a microscopic real-space interpretation of the head-to-tail arrangement.

\section{Conclusion}\label{sec_conclusion}
We have investigated shift-current generation in single-domain and multidomain
2D SSH systems by combining symmetry analysis with numerical calculations.
The head-to-head and head-to-tail arrangements act as symmetry-selective
elements, and the mutually exclusive sets of tensor components they admit are
fixed by the symmetry of the full multidomain supercell. For the supercells
studied in the main text, the two geometries differ not only in which direction is
selected but also in the magnitude of the
surviving component, which stays close to the single-domain value for the
head-to-head arrangement and is roughly seven times smaller for the
head-to-tail one. For these supercells, equal-area averages of the
constituent single-domain spectra do not reproduce the detailed multidomain spectra, and a bond-resolved
analysis clarifies the real-space origin of this discrepancy. The multidomain
response is therefore a property of the full geometry rather than a simple sum
of independent single-domain responses.
Appendix~\ref{sec:appendixC} extends each domain to five and ten primitive cells along $x$. The same
symmetry selection applies, while the allowed spectra depend on the domain
width and the representative bond contributions extend into the domain
interiors. Over this range of widths the dominant head-to-tail response
changes sign and grows by about an order of magnitude, exceeding the
head-to-head response. Thus, the ordering of the two magnitudes found at the
smallest supercell does not carry over to wider domains. All these calculations describe periodic interface pairs rather
than an isolated domain wall.

The scope of the present model has three limitations. First, the model describes
prescribed polar configurations. Their microscopic formation and switching
are outside its scope, as discussed in Sec.~\ref{sec:model-polarization}.
Second, the width extension in Appendix~\ref{sec:appendixC} samples finite
periodic supercells. The calculation does not establish an isolated-wall limit or separate
an independent bulk response from a wall contribution. The symmetry-imposed
vanishing of the forbidden tensor components is exact for widths that
preserve the same mirror or glide symmetry, whereas the magnitudes and line
shapes of the allowed components depend on the superlattice period. Third, the bond-resolved
decomposition is basis dependent, and only the summed conductivity is
observable.

\begin{figure}
  \centering
  \includegraphics[width=85mm]{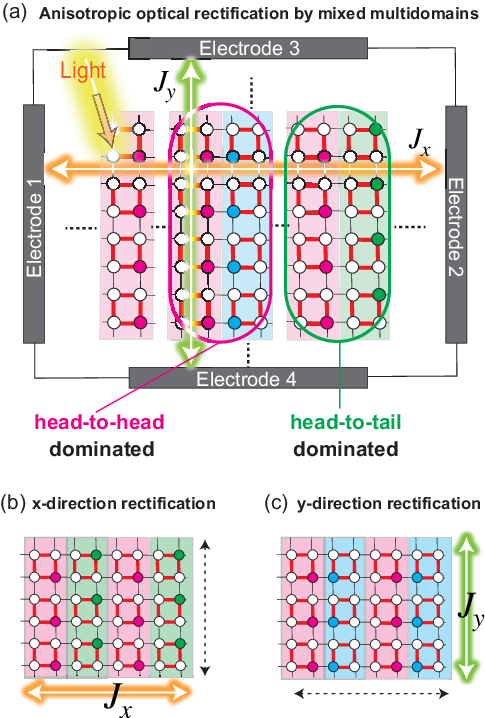}
  \caption{
    Conceptual illustration of anisotropic optical rectification in a mixed multidomain structure.
(a) Conceptual arrangement containing head-to-head-dominated and head-to-tail-dominated multidomain regions under light irradiation.
The head-to-tail-dominated region mainly contributes to the photocurrent along the $x$ direction between electrodes 1 and 2, whereas the head-to-head-dominated region mainly contributes to the photocurrent along the $y$ direction between electrodes 3 and 4.
(b) Schematic representation of $x$-direction rectification in a head-to-tail-dominated region.
(c) Schematic representation of $y$-direction rectification in a head-to-head-dominated region.
Here, the head-to-head-dominated and head-to-tail-dominated regions correspond to the periodic structures defined in Sec.~\ref{sec:model-polarization}.
The indicated response directions are motivated by the dominant conductivity components calculated for the two multidomain structures and do not represent a device-level transport calculation.
  }
  \label{fig:device_concept}
\end{figure}
Figure~\ref{fig:device_concept}(a) illustrates the resulting design principle for anisotropic optical rectification.
Head-to-head-dominated and head-to-tail-dominated regions may preferentially direct responses along the $y$ and $x$ directions, as sketched in Figs.~\ref{fig:device_concept}(c) and (b), respectively. Because the two geometries are built from identical polar building blocks and differ only in how those blocks are joined, the arrangement of prescribed polar configurations may serve as a design parameter for selecting a preferred rectification direction. This concept is motivated by the dominant conductivity components of the present bulk-periodic model, and it does not constitute a device-level transport calculation including electrodes, contacts, or carrier collection.

The experimentally observed dependence of the photovoltaic response on the
domain-wall arrangement in SnS~\cite{Nanae2026DomainSnS} motivates examining
symmetry constraints in addition to transport and local wall conductivity.
Applying the present selection rules to a specific SnS domain structure
requires identifying the symmetries of that structure.
Within the present model, the domain pattern alone restricts which
tensor components can contribute for a given illumination and current direction,
before any transport considerations enter. A quantitative comparison with
experiment would nevertheless require a material-specific electronic
structure, realistic domain widths, and a device-level transport
calculation, none of which is attempted here.

Although minimal, the present 2D SSH model captures how the symmetry and geometry of domain structures can tailor the
tensor components and the preferred response direction of shift current. This
provides a compact theoretical framework for domain-engineered nonlinear
optical responses in polar multidomain systems, and a symmetry-based starting
point for interpreting orientation-dependent photocurrents in real multidomain
ferroelectrics such as SnS~\cite{Nanae2026DomainSnS}.

\section*{ACKNOWLEDGMENTS}
This work was supported by JSPS KAKENHI (Grant Nos. JP25K01609, JP22H05473, and JP21H01019) and JST CREST (Grant No. JPMJCR19T1).
T.~K. acknowledges support from the Sasakawa Scientific Research Grant from The Japan Science Society and the Yashima Environment Technology Foundation.
K.~W. acknowledges financial support from the Sumitomo Foundation for Basic Science Research Projects (Grant No. 2401203).

\section*{Data Availability}
The data that support the findings of this study are available from the corresponding author upon reasonable request.

\appendix
\section{Other Components of Shift-Current Conductivity}
\label{sec:appendixA}
Additional tensor components for the head-to-head arrangement complement the main-text discussion of the dominant components.
\begin{figure}
  \centering
  \includegraphics[width=85mm]{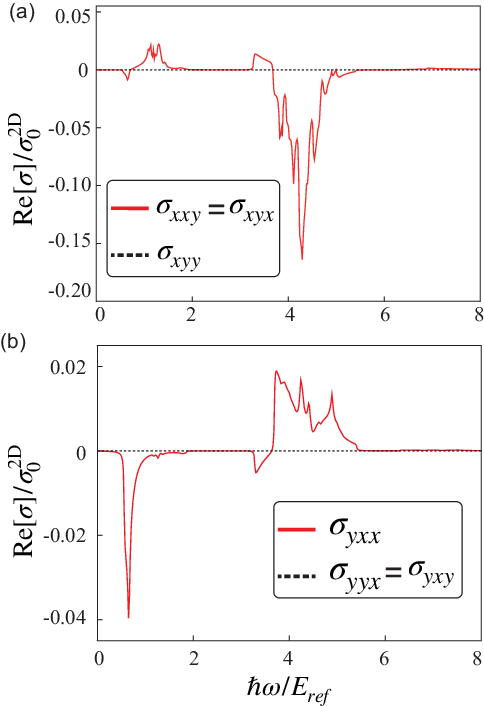}
  \caption{Real parts of (a) $\sigma_{xxy}$, $\sigma_{xyx}$, and $\sigma_{xyy}$, and
(b) $\sigma_{yxx}$, $\sigma_{yxy}$, and $\sigma_{yyx}$, for the head-to-head arrangement.
Red solid lines show the symmetry-allowed components, namely
$\sigma_{xxy}=\sigma_{xyx}$ in (a) and $\sigma_{yxx}$ in (b), while black dotted
lines show the symmetry-forbidden ones, namely $\sigma_{xyy}$ in (a) and
$\sigma_{yxy}=\sigma_{yyx}$ in (b), which vanish to numerical accuracy.
The conductivity is reported as $\mathrm{Re}[\sigma_{ijk}]/\sigma_0^{\rm 2D}$.}
  \label{fig:other_conductivity}
\end{figure}
Figure~\ref{fig:other_conductivity}(a) shows $\sigma_{xxy}$, $\sigma_{xyx}$, and $\sigma_{xyy}$, and Fig.~\ref{fig:other_conductivity}(b) shows $\sigma_{yxx}$, $\sigma_{yxy}$, and $\sigma_{yyx}$, for the head-to-head arrangement.
The components allowed by $\hat{M}_x$, namely $\sigma_{xxy}$, $\sigma_{xyx}$,
and $\sigma_{yxx}$, are numerically finite, whereas the forbidden
$\sigma_{xyy}$, $\sigma_{yxy}$, and $\sigma_{yyx}$ vanish to numerical
accuracy across the entire photon-energy range. The same holds for the
forbidden $\sigma_{xxx}$ in
Fig.~\ref{shift_current_multi_domain_headtohead}(b). The selection rule
derived in Sec.~\ref{sec:symmetry-constraints} is thus reproduced by the
calculation without being imposed on it, which also serves as an internal
consistency check of the numerical implementation. The equality
$\sigma_{xxy}=\sigma_{xyx}$ follows from permutation of the two optical input
indices in the symmetrized conductivity~\cite{AversaSipe1995,SipeShkrebtii2000},
not from $\hat{M}_x$ itself.
The finite $\sigma_{xxy}=\sigma_{xyx}$ generates the
$x$-directed response of the head-to-head arrangement under obliquely
polarized light, as discussed at the end of
Sec.~\ref{sec:symmetry-constraints}.
The head-to-tail arrangement admits the complementary set $\sigma_{xyy}$,
$\sigma_{yxy}$, $\sigma_{yyx}$, and $\sigma_{xxx}$, obtained from the
head-to-head set by exchanging the roles of $x$ and $y$ in the selection rule.
Here we analyze $\sigma_{xxx}$, which is the component
excited by illumination polarized along $x$. Among the remaining components,
$\sigma_{xyy}$ contributes for $y$-polarized illumination, whereas
$\sigma_{yxy}$ and $\sigma_{yyx}$ contribute when the incident polarization
contains both $x$ and $y$ components.

\section{Polarization-Angle Dependence of Shift-Current Conductivity}
\label{sec:appendixB}
We relate the polarization-angle dependence of the shift-current conductivity to the underlying crystal symmetries.

For linearly polarized light incident normal to the 2D plane,
we define the in-plane polarization unit vector as
$\mathbf{e}(\phi)=(\cos\phi,\sin\phi)$,
where $\phi$ is measured from the $x$ axis.
The optical field is then given by
\[
E_j(\omega)=E_0 e_j(\phi),
\]
where $E_0$ is the amplitude of the electric field and $e_j(\phi)$ is the $j$th ($j=x,y$) component of $\mathbf{e}(\phi)$.
The polarization-angle-dependent shift-current sheet density is defined by
\[
J_i^{\rm shift}(\phi;\omega)
=
2E_0^2\sigma_i^{\rm eff}(\phi;\omega),
\]
with
\[
\sigma_i^{\rm eff}(\phi;\omega)
=
\sum_{j,k=x,y}
\sigma_{ijk}(0;\,\omega,-\omega)e_j(\phi)e_k(\phi).
\]
Equivalently,
\begin{align*}
\sigma_i^{\rm eff}(\phi;\omega)
= {}& \sigma_{ixx}\cos^2\phi
+(\sigma_{ixy}+\sigma_{iyx})\sin\phi\cos\phi \\
&+\sigma_{iyy}\sin^2\phi.
\end{align*}
Thus, the polarization-angle dependence is determined not only by the diagonal components
\(\sigma_{ixx}\) and \(\sigma_{iyy}\), but also by the mixed components
\(\sigma_{ixy}\) and \(\sigma_{iyx}\).

\begin{figure}
  \centering
  \includegraphics[width=85mm]{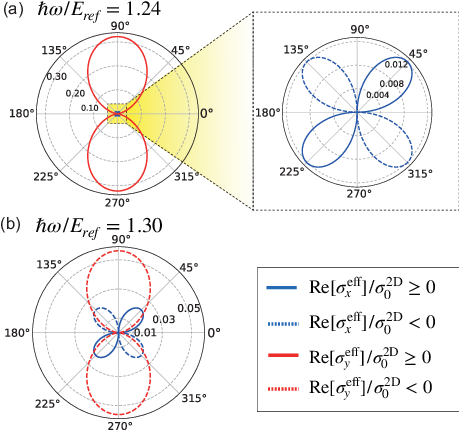}
  \caption{Polarization-angle dependence of the real parts of the effective
shift-current conductivities $\sigma_x^{\rm eff}$ and $\sigma_y^{\rm eff}$ for
the head-to-head arrangement at
(a) $\hbar\omega/E_{\rm ref}=1.24$ and
(b) $\hbar\omega/E_{\rm ref}=1.30$.
Blue and red curves denote the $x$- and $y$-direction responses, respectively.
The radial distance represents $|\mathrm{Re}[\sigma_i^{\rm eff}]|/\sigma_0^{\rm 2D}$,
and solid and broken curves distinguish nonnegative from negative values, as
indicated in the legend.
The inset in (a) enlarges the small $\sigma_x^{\rm eff}$ component and uses its
own radial scale.
}
\label{fig:polarization_dependence}
\end{figure}
The photon energies in Fig.~\ref{fig:polarization_dependence} correspond to the characteristic energies indicated in Fig.~\ref{fig:localized_current}(a).
At $\hbar\omega/E_{\rm ref}=1.24$ [Fig.~\ref{fig:polarization_dependence}(a)], the $y$-direction response $\sigma_y^{\rm eff}$ is larger than the $x$-direction response $\sigma_x^{\rm eff}$ over a wide range of polarization angles. This behavior is consistent with the large positive $\mathrm{Re}[\sigma_{yyy}]$ and the vanishing $\mathrm{Re}[\sigma_{xxx}]$ seen in Fig.~\ref{shift_current_multi_domain_headtohead}(b).
The polarization-angle dependence of $\sigma_y^{\rm eff}$ is approximately proportional to $\sin^2\phi$, reflecting the dominance of the $\sigma_{yyy}$ component.
In contrast, $\sigma_x^{\rm eff}$ is approximately proportional to $\sin\phi\cos\phi$, but its magnitude is much smaller than that of $\sigma_y^{\rm eff}$, indicating that the $x$-direction response is strongly suppressed at this photon energy.
At $\hbar\omega/E_{\rm ref}=1.30$ [Fig.~\ref{fig:polarization_dependence}(b)], $\sigma_y^{\rm eff}$ changes sign and becomes negative, while $\sigma_x^{\rm eff}$ remains much smaller in magnitude. The angular dependence of $\sigma_y^{\rm eff}$ still approximately follows $\sin^2\phi$, indicating that the sign reversal originates from $\mathrm{Re}[\sigma_{yyy}]$ rather than from a change in the dominant polarization channel. By contrast, $\sigma_x^{\rm eff}$ exhibits a $\sin\phi\cos\phi$-like dependence at both photon energies, indicating that the mixed components $\sigma_{xxy}$ and $\sigma_{xyx}$ determine the angular profile of the subdominant $x$-direction response.
Thus, the dominant polarization-angle-dependent shift-current response is controlled by the $\sigma_{yyy}$ channel, whereas the mixed tensor components mainly contribute to a weaker $x$-direction response.

\section{Domain-Width Dependence of the Shift-Current Response}
\label{sec:appendixC}
We define $N$ as the number of primitive cells along $x$ in each domain,
so that each domain has width $Na$. The supercell holds two
equally wide domains, with length $2Na$, $2N$ primitive cells, and $8N$
sites. We study $N=1$ in the main text (Figs.~\ref{fig:2d-ssh-model_DW}
and~\ref{fig:2d-ssh-multi-energy_band}). Here we extend the arrangements to
$N=5$ and $10$
(40 and 80 sites), retaining the hopping amplitudes and on-site potentials of
the main text.
The central and periodic interfaces lie at $x/a=N$ and 0, and the domain centers at $N/2$ and $3N/2$. Periodicity identifies $x/a=0$ and $2N$ as the same interface.
This extension preserves the mirror symmetry of the head-to-head arrangement and the glide symmetry of the head-to-tail arrangement. The selection rules of Sec.~\ref{sec:symmetry-constraints} therefore remain valid. 

\begin{figure*}[tbp]
\centering
\includegraphics[width=170mm]{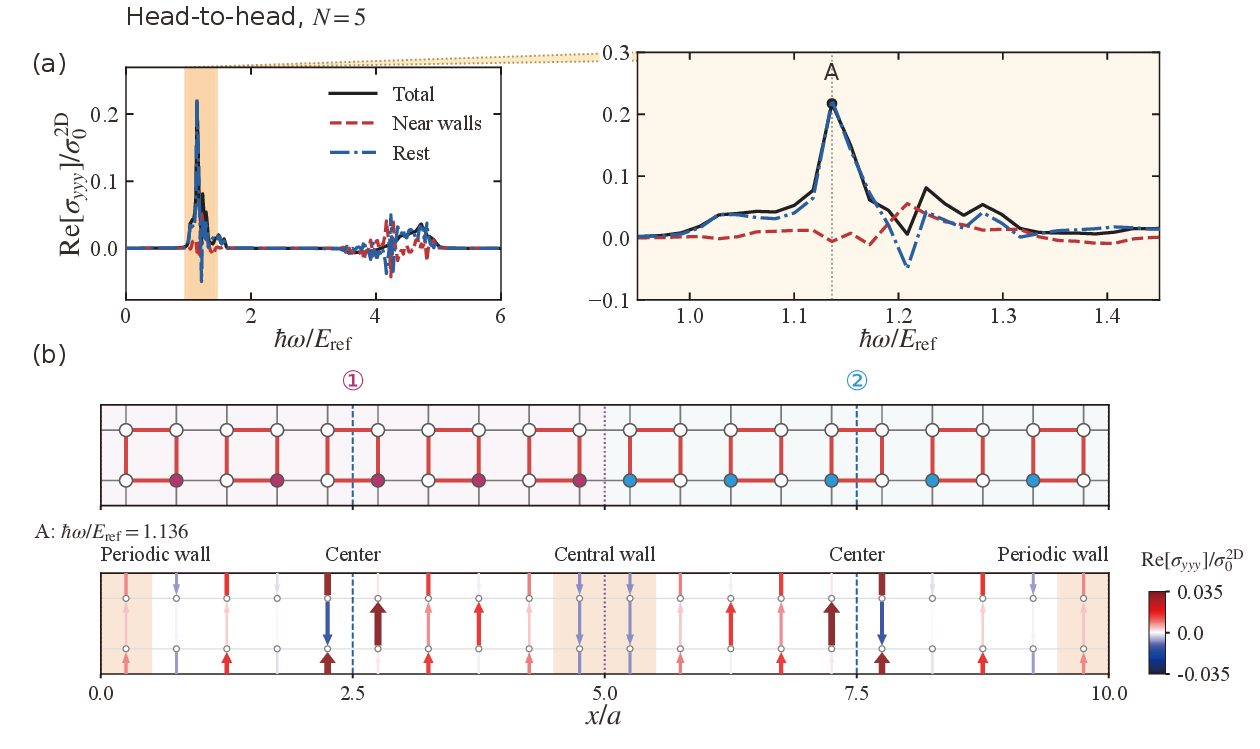}
\caption{Head-to-head supercell with $N=5$. (a) Total (black solid), near walls (red dashed), and rest (blue dash-dotted) spectra. The left panel shows the overview; the right panel enlarges the shaded frequency interval. (b) Schematic domain structure (upper panel) and spatial map of the bond-resolved shift-current contributions at A, $\hbar\omega/E_{\rm ref}=1.13627$ (lower panel).}
\label{fig:width_hh5}
\end{figure*}

\begin{figure*}[tbp]
  \centering
\includegraphics[width=170mm]{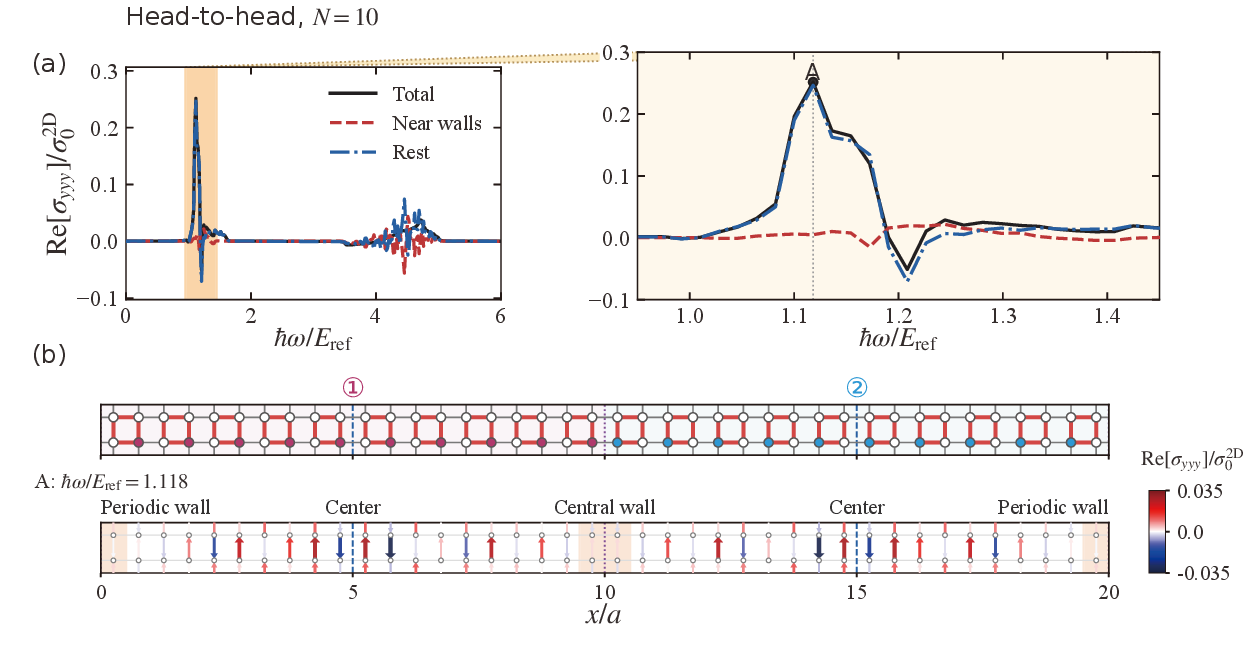}
\caption{Head-to-head supercell with $N=10$. (a) Total (black solid), near walls (red dashed), and rest (blue dash-dotted) spectra. The left panel shows the overview; the right panel enlarges the shaded frequency interval. (b) Schematic domain structure (upper panel) and spatial map of the bond-resolved shift-current contributions at A, $\hbar\omega/E_{\rm ref}=1.11824$ (lower panel).}
\label{fig:width_hh10}
\end{figure*}

\begin{figure*}[tbp]
\centering
\includegraphics[width=170mm]{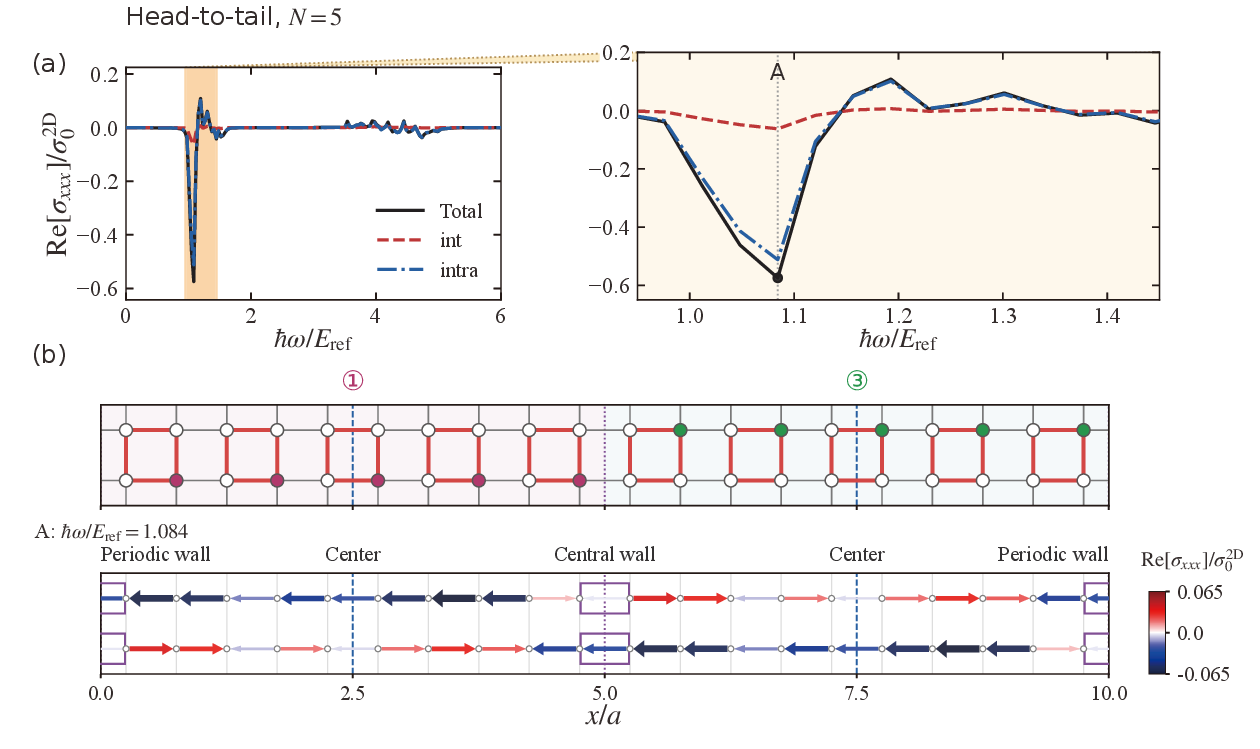}
\caption{Head-to-tail supercell with $N=5$. (a) Total (black solid), interface-crossing (int, red dashed), and intradomain (intra, blue dash-dotted) spectra. The left panel shows the overview; the right panel enlarges the shaded frequency interval. (b) Schematic domain structure (upper panel) and spatial map of the bond-resolved shift-current contributions at A, $\hbar\omega/E_{\rm ref}=1.08434$ (lower panel).}
\label{fig:width_ht5}
\end{figure*}

\begin{figure*}[tbp]
\centering
\includegraphics[width=170mm]{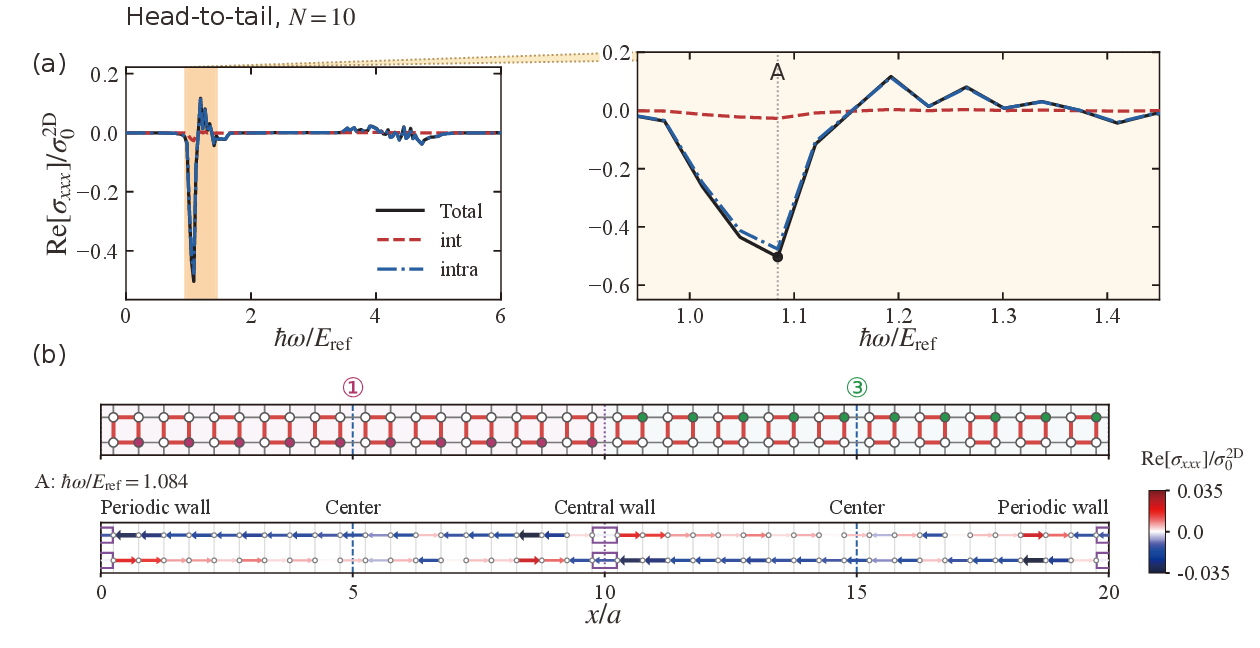}
\caption{Head-to-tail supercell with $N=10$. (a) Total (black solid), interface-crossing (int, red dashed), and intradomain (intra, blue dash-dotted) spectra. The left panel shows the overview; the right panel enlarges the shaded frequency interval. (b) Schematic domain structure (upper panel) and spatial map of the bond-resolved shift-current contributions at A, $\hbar\omega/E_{\rm ref}=1.08434$ (lower panel).}
\label{fig:width_ht10}
\end{figure*}

Figures~\ref{fig:width_hh5}--\ref{fig:width_ht10} show the spectra, domain
structures, and bond-contribution maps for both arrangements at $N=5$ and
$10$. We retain $\eta/E_{\rm ref}=0.01$, the conductivity unit
$\sigma_0^{\rm 2D}$, the unit-area normalization for supercell area $2Na^2$,
and the photon-energy sampling of the main text: over
$0\leq\hbar\omega/E_{\rm ref}\leq9$, 500 points with
$\Delta(\hbar\omega/E_{\rm ref})\simeq0.0180$ for the head-to-head
supercells and 250 points with
$\Delta(\hbar\omega/E_{\rm ref})\simeq0.0361$ for the head-to-tail
supercells. The Brillouin-zone integrals use periodic midpoint meshes: along
$\ell=x,y$ the sampled wave numbers are uniformly spaced and offset by half a
mesh spacing from $-\pi/L_{\ell}$, with $L_x=2Na$ and $L_y=a$. Refining the
meshes from $800\times800$ to $1200\times1200$ (head-to-head) and from
$200\times200$ to $400\times400$ (head-to-tail) changes the dominant
extremum by less than $0.4\%$. Individual
bond contributions at that photon energy change by up to about $5\%$ of the
largest one. Thus, the total response is better converged than its
bond-resolved decomposition.

The circled labels in the structural panels identify the single-domain
configurations of Fig.~\ref{fig:single-domain-polarization}, and blue dashed
lines mark domain centers. Arrow direction gives the sign along Cartesian
$y$ (head-to-head) or $x$ (head-to-tail), while color and thickness encode
magnitude. The color scales differ between figures. Quantitative comparisons need the color bars.

\noindent\textit{Head-to-head.}
Figures~\ref{fig:width_hh5}(a) and~\ref{fig:width_hh10}(a) show $\sigma_{yyy}$ for
$N=5$ and $10$, respectively, calculated on $1200\times1200$ periodic
midpoint $\mathbf{k}$ meshes. The response remains finite at both widths.
On the photon-energy grid used here, the sampled peak lies at
$\hbar\omega/E_{\rm ref}\simeq1.24$ at $N=1$ and at 1.14 and 1.12 at
$N=5$ and $10$, respectively. The shift from $N=1$ spans about six grid
spacings, whereas the values at $N=5$ and $N=10$ differ by one spacing and
the grid leaves that difference unresolved. The peak height remains of the
same order of magnitude at all three widths.

Only vertical bonds contribute to the bond-resolved $y$-directed response
(Sec.~\ref{sec:real-space}). The curve labeled ``near walls'' sums the
contributions of bonds whose midpoints lie within the shaded regions of width
$a$ centered on the two interfaces, and the ``rest'' curve holds the
remainder. A midpoint exactly on a boundary contributes equally to both
curves.
Figures~\ref{fig:width_hh5}(b) and~\ref{fig:width_hh10}(b) map the bond
contributions at A, the main peak for each width. The contributions extend
into both domain interiors. The signed near-wall sum is a small part of
the total, although cancellation can conceal larger individual contributions.

\noindent\textit{Head-to-tail.}
Figures~\ref{fig:width_ht5}(a) and~\ref{fig:width_ht10}(a) show $\sigma_{xxx}$ for
$N=5$ and $10$, respectively, calculated on $400\times400$ periodic
midpoint $\mathbf{k}$ meshes. The spectra have a dominant negative dip near
$\hbar\omega/E_{\rm ref}=1.08$, whereas the main peak at $N=1$ is positive
and lies near 1.23. The dominant extremum of
$\mathrm{Re}[\sigma_{xxx}]/\sigma_0^{\rm 2D}$ is $+0.047$ at $N=1$,
$-0.574$ at $N=5$, and $-0.503$ at $N=10$. Widening the domains therefore
changes the sign of the dominant response and raises its magnitude by about
an order of magnitude. The corresponding head-to-head peak is $0.252$ at
$N=10$. Thus, the head-to-tail response, about seven times smaller than the
head-to-head one at $N=1$, is about twice as large at $N=10$. These are the extrema sampled on the
photon-energy grid used here, at the two widths studied.

Following Sec.~\ref{sec:real-space}, the interface-crossing contribution
$\sigma_{xxx}^{\rm int}$ sums the four bonds outlined in purple, and the
intradomain contribution $\sigma_{xxx}^{\rm intra}$ sums the rest. Figures~\ref{fig:width_ht5}(b)
and~\ref{fig:width_ht10}(b) map these contributions at the common frequency
A, $\hbar\omega/E_{\rm ref}\simeq1.084$. The contributions extend into both
domain interiors. Both sums are negative at $N=5$ and $10$, with the
intradomain sum accounting for most of the total, whereas at $N=1$ they have opposite
signs and cancel in part at the same frequency.
Thus the selection rules hold at both widths, whereas the spectra depend on
domain width.

\onecolumngrid
\clearpage
\twocolumngrid
\bibliography{refs}
\end{document}